\documentclass[11pt]{article}
\usepackage{amsmath,amssymb,amsthm}
\usepackage{enumitem}
\usepackage{hyperref}
\usepackage{pgfplots}
\usepackage{graphicx}
\usepackage{array}
\usepackage{multirow}
\usepackage{authblk}
\usepackage{wrapfig}
\usepackage[margin=1in]{geometry}
\usepackage{placeins}
\pgfplotsset{compat=1.17}
\usepackage[numbers,sort&compress]{natbib}

\newcommand\ap[1]{\textcolor{black}{#1}}

\begin{document}

\title{Predictive capabilities of \\
the integrated modeling TRANSP code \\
for tokamak plasmas}
\author[1]{A.Y. Pankin}
\author[1]{J. Breslau}
\author[1]{M.V. Gorelenkova}
\author[1]{R. Budny}
\author[2]{M. Goliyad}
\author[3]{B.A. Grierson}
\author[1]{G.W. Hammett}
\author[1]{S.C. Jardin}
\author[3,4]{J.B. Lestz}
\author[5]{X. Yuan}
\affil[1]{Princeton Plasma Physics Laboratory, Princeton, NJ 08540}
\affil[2]{Rutgers University, Piscataway, NJ 08854}
\affil[3]{General Atomics, San Diego, CA 92121}
\affil[4]{University of California, Irvine, CA 92697}
\affil[5]{Argonne National Laboratory, Lemont, IL 60439}
\date{}
\maketitle

\newcolumntype{C}[1]{>{\centering\arraybackslash}p{#1}}  

\begin{abstract}
This paper expands on the TRANSP description given in \textit{Computer Physics Communications 312 (2025) 109611} by describing recent progress in TRANSP's predictive functionality and emphasizing development of the PT\_SOLVER module and integration of the high-fidelity T3D/GX framework for plasma profile prediction using a high-fidelity gyrokinetic model for turbulent transport. PT\_SOLVER is a modular, multi-region, parallel solver for coupled transport equations of particle density, electron and ion energy, and toroidal angular momentum that uses an implicit Newton method to advance the solution of these equations. The numerical formulation includes source coupling, moving-geometry terms, and nonlinear stabilization based on modified P\'{e}clet numbers, thereby enabling the PT\_SOLVER to handle the stiffness associated with gradient-dependent transport models. Stabilization occurs via a nonlinear function controlling discretization in zones of steep gradients or rapidly changing transport coefficients. Source terms that account for heating, current drive, alpha-particle effects, and collisional energy exchange are handled thoroughly, and both residual norms and profile-change measures are used to assess convergence. Verification is carried out using analytical benchmark solutions, manufactured solution benchmarks, convergence studies of stiff gradient-dependent diffusivities, and code-to-code comparisons of TGYRO using the TGLF/NEO models for anomalous and neoclassical transport. A multi-level parallel decomposition strategy in PT\_SOLVER is described for both flux surface and TGLF wavenumber parallelism, along with representative scaling results. This paper also describes the TRANSP Interface to the modular T3D/GX workflow and presents verification examples related to the interface for coupled prediction simulations. Agreement between PT\_SOLVER and the modular T3D solver is confirmed by a benchmark set under rising complexity in physics. The results in this paper confirm that the predictive TRANSP framework has a robust numerical implementation for time-dependent predictive transport simulations, and it provides a basis for future hybrid reduced and high-fidelity workflows.
\end{abstract}


\section{Introduction}
The development of controlled thermonuclear fusion offers a pathway to clean and sustainable energy. Tokamaks are among the most promising magnetic confinement devices, and the TRANSP integrated modeling code~\cite{RHawryluk1980,JBreslau2018,APankin2024} has long served as a critical tool for both interpretive and predictive analysis of tokamak plasmas. For predictive analysis, the code is actively used for model verification and validation, plasma scenario planning and assessments of future tokamak devices~\cite{Budny_2012,Kim_2023,BudnyCordey2016,BudnyJET2018,Abbate2024,Avdeeva_2023,Lestz2025_AssessingNSTX,Lestz2025_MMMSensitivityNSTX,Auriemma_2026}.

Integrated predictive modeling plays an important role in tokamak fusion research. It is essential for optimizing experiments, developing scenarios, and determining how a new regime will operate. Predictive modeling can provide a series of evolving plasma profiles in time consistent with realistic sources of heating, particles, and momentum; transport of particles and energy; and plasma equilibrium. Interpretive analysis results are being used to create databases of plasma discharges. Predictive simulations are used to evaluate the effects of changing the heating, fueling, current drive, rotation, impurity level or control paths \textit{prior} to performing an experiment. They are used for designing future tokamaks such as ITER. The simulation will only be valid if the underlying physics-based transport models used for the computations have a high level of physics fidelity and a robust numerical implementation.

TRANSP's long history as a leading integrated modeling tool for tokamak analysis stems from its use in interpretive analyses of the experimental data through the combination of data from experimental measurements, equilibrium reconstructions, and source calculations, which are combined to infer transport and power balance. The code can also provide predictive capability and time evolution of plasma profiles through the same integrated framework using theory-based transport models and consistent treatments of heating, current drive, particle, and momentum sources. The close connection between interpretive and predictive functions in TRANSP is one of its key strengths, as both rely on the same infrastructure to provide geometries, evaluate sources, and couple physics. This establishes the direct applicability of predictive calculations to experimental analyses and validations of models.

TRANSP simulates the time-dependent evolution of plasma profiles, including density, temperature, and toroidal rotation, by integrating coupled conservation equations that account for heating, current drive, and complex equilibrium effects. In this paper, ``predictive'' TRANSP refers specifically to simulations in which the time evolution of kinetic profiles, such as temperature, density, and toroidal rotation, is advanced using transport equations and theory-based models for anomalous and neoclassical transport. This is distinct from other time-dependent TRANSP capabilities, such as current-diffusion workflows, which may also evolve plasma quantities in time but are not the focus of the present work. 

Other integrated modeling workflows commonly used for transport prediction include ASTRA~\cite{Pereverzev2002_ASTRA}, JETTO~\cite{Cenacchi1988_JETTO}, JINTRAC~\cite{Romanelli2014_JINTRAC}, and IPS-FASTRAN~\cite{Park2017,Park2018_IPSFASTRAN}. These tools address related predictive-transport problems, but they differ from TRANSP/PT\_SOLVER in workflow organization, coupling strategy, and the degree to which source, equilibrium, and transport physics are embedded within a single established TRANSP framework. Recent workflows relevant to predictive transport also include FUSE~\cite{Meneghini2024_FUSE}, which supports integrated steady-state and time-dependent plasma modeling in a broader fusion-design framework, and TORAX~\cite{Citrin2024_TORAX}, a fast differentiable core transport simulator implemented in JAX.

A key module within predictive TRANSP is PT\_SOLVER, which integrates the nonlinear transport equations using an implicit, Newton-based iterative method. Convergence is monitored by a normalized residual, combined with maximum relative change and maximum pointwise residual criteria. 

A detailed overview of the capabilities of the code, including the general predictive process and the options available for prediction, has recently been presented in Ref.~\cite{APankin2024}. Therefore, the purpose of this article is not to duplicate this overview, but to deal with one of the important elements of the code that was not described with sufficient algorithmic detail there. 

Three important features of PT\_SOLVER are worth mentioning here. First, PT\_SOLVER uses a multi-region approach whereby different anomalous and neoclassical transport models may be used in different radial regions, such as axial, confinement, and edge regions. This provides useful flexibility in cases where a single transport model may not be sufficient to describe the physics throughout the whole radius of the plasma -- for example, when an anomalous or neoclassical model is not valid in a particular plasma region. In particular, it might not be valid near the magnetic axis or in the plasma edge. The original reference TRANSP paper~\cite{APankin2024} includes the description of how the transport models can be combined for different confinement regions. Second, PT\_SOLVER was designed for multilevel parallelization to ensure that both the transport advance and the computational expense of evaluating the transport models could be distributed using modern computing architectures. Specifically, the implementation of the anomalous transport TGLF model in PT\_SOLVER includes parallelism over the flux surfaces combined with parallelism over the TGLF wavenumbers, thus eliminating the constraint wherein the number of processors would have to equal the number of transport zones. Lastly, PT\_SOLVER is engineered to address the stiffness associated with the gradient sensitivity of transport closures in a turbulence model by implementing an implicit Newton-based nonlinear solver enhanced with stabilization techniques based on the modified P\'{e}clet number.

Previous work by Jardin \textit{et al}.~\cite{jardin08} identified the numerical difficulties related to the use of gradient-dependent transport coefficients, emphasized the need to develop robust stabilization techniques, and provided an outline of a numerical scheme to address these difficulties.  The code's current implementation of the predictive process in PT\_SOLVER differs from this numerical scheme, and there is no documented reference on the current PT\_SOLVER numerical algorithm. Indeed, the implicit multi-channel method used by TRANSP has specific features related to source treatment, moving geometry, P\'{e}clet number stabilization, and nonlinear convergence tests. Therefore, a detailed description of the code's predictive process is provided as a reference for users of the code and as a basis for code-to-code verification of the predictive solver. This description should assist in clearly identifying both the capabilities and current limitations associated with predictive transport simulations performed with TRANSP.

The primary motivation for focusing on this area is due to the computational intensity of predictive transport simulations in tokamaks, even though their established transport models are considered relatively mature. The equations governing transport are highly nonlinear, tightly coupled over multiple transport channels, and will be sensitive to geometry, boundary conditions, collisional interactions, auxiliary heating, particle sources, and momentum input. Additionally, several transport closures used in predictive studies generally yield diffusion and pinch coefficients that depend greatly on local gradients and also can change drastically over space and time. Therefore, in this environment, it is very difficult to use explicit approaches or naive implicit discretizations, which often lead to imprecise, oscillatory, and/or unstable solutions. Thus, a successful predictive working process requires an appropriate transport model and a transport solver that provides stable, modular, and reliable convergence behavior across a wide range of plasma operation conditions.

PT\_SOLVER is the component in TRANSP designed to address these requirements.  PT\_SOLVER provides a multi-channel, implicit transport solution for predicting how the temperatures of electrons and ions, plasma density, composition of impurities, and toroidally rotating plasma evolve over time using the standard TRANSP flux-surface average framework. PT\_SOLVER includes a selection of anomalous and neoclassical transport models that are coupled to the rest of TRANSP's integrated modeling workflow. Since PT\_SOLVER shares the same TRANSP framework that is used for calculations of equilibrium and sources, it is internally self-consistent and can be applied for full-device predictive simulations. For this reason, in order to provide confidence in the interpretation of predicted results from TRANSP, it is essential to have a clearly defined numerical description, along with an appropriate verification strategy for PT\_SOLVER.

The verification method employed in this paper focuses on the verification of PT\_SOLVER itself, not the models for anomalous and neoclassical transport that it can use. We do not compare the results with experiment, but only with the predictions of other solvers that utilize the same theory-based model for anomalous transport. The objective here is to verify that the prediction from PT\_SOLVER yields the same profiles as those predicted by a different solver, given the same inputs. The focus of this verification exercise is the coupling or numerical implementation. This type of comparison also serves as a complement to internal testing against simplified cases and provides a higher quality of information than simply looking at profile comparisons for the two cases.


Another aspect related to the solver-centric focus of the TRANSP code is its recent extension to allow more modular approaches to transport and turbulence workflows, such as coupling to T3D~\cite{Qian22,Barnes2008} and the nonlinear gyrokinetic GX model~\cite{Mandell2018,Mandell2024}. This is significant because it supports the development of predictive calculations by coupling profile evolutions to higher-fidelity turbulence models, including GPU-based nonlinear gyrokinetic calculations. However, the goal of the present paper is not to provide a comprehensive verification and performance assessment of the coupling of the TRANSP/T3D/GX workflow, but rather to give an overview of the architecture of the coupling, show stable operation in an integrated environment, and to provide examples of how the predictive TRANSP framework can be connected to external transport and turbulence components. 

The core contributions of this work are as follows. First, we present a comprehensive overview of the PT\_SOLVER formulation used for predictive TRANSP simulations, providing detailed descriptions of the transport equations, discretization methods, nonlinear solution strategies, stabilization methods, and convergence criteria. Second, we describe the verification efforts for PT\_SOLVER using a series of benchmark cases with the TGYRO code. Finally, we give an overview of the recent integration development between TRANSP, T3D, and GX that illustrates a modular approach to the extension of the predictive framework. By providing detailed information on the current development of predictive workflows, we have provided a primary methodological and verification contribution relating specifically to the predictive transport functionality of the TRANSP framework.

The 2025 TRANSP technical reference paper~\cite{APankin2024} provided general organization of the TRANSP predictive workflow, a summary of the capability of the TRANSP predictive code, and an overview of the transport modeling and predictive tools available for TRANSP. The emphasis of this paper is to present three specific contributions related to the PT\_SOLVER numerical formulation. The first contribution is a detailed description of this formulation, including the discretization method used by PT\_SOLVER, the use of coupling of the source terms, the method of handling moving geometry, the method of stabilizing nonlinearity using an adjusted P\'{e}clet number criterion, and the criteria that are used for convergence of iterative solutions in practical applications of the solver. The second contribution is to present a comprehensive hierarchy of verification tests for the predictive portion of the TRANSP code. This hierarchy includes analytically solved and manufactured solution verification tests, convergence tests for stiff gradient-dependent transport coefficients, and a code-to-code verification for PT\_SOLVER and TGYRO using TGLF and NEO models for anomalous and neoclassical transport. The third contribution is to document key implementation features that were not discussed in detail in the above-referenced papers. These features include the multilevel parallelization method used in PT\_SOLVER calculations for transport resolution and the initial modular coupling of the TRANSP to the T3D/GX predictive calculations. In this sense, the present paper is intended as a methodological and verification-focused supplement to the broader TRANSP capabilities paper, rather than as a second general review of predictive functionality.

This paper is structured as follows: Section 2 includes background and motivation. Section 3 describes the governing transport equations, predictive transport framework, and assumptions made in the flux surface averaged formulation. In Section 4, we introduce the PT\_SOLVER numerical algorithm, which incorporates the implicit Newton formulation, coefficient construction, stabilization procedure, and convergence metrics. It also includes additional cross-verification studies between PT\_SOLVER and TGYRO. Section 5 discusses the integration path with T3D/GX and provides a first demonstration of the coupled workflow. Finally, Section 6 summarizes the current capabilities, limitations, and future direction of this work.

\section{Background and motivation}
\label{sec:background}

Predictive transport simulations are an essential part of the integrated modeling process for a tokamak. The ability to calculate how the plasma profiles evolve over time self-consistently under the influence of prescribed sources, sinks and boundary conditions is a critical part of scenario development, experiment planning, actuator optimization, and extrapolation to future devices. In practical integrated modeling, this problem is typically formulated as a coupled set of one-dimensional transport equations written on magnetic flux surfaces and coupled to the time evolution of the equilibrium, heating and current drive, particle fueling, momentum input, and models for large-scale MHD events, such as sawtooth crashes. The importance of the overall integrated modeling context has been acknowledged for many years. The dynamic modeling of tokamak transport together with plasma position and control has been demonstrated previously~\cite{Jardin1986} and described in detail in the literature~\cite{Jardin2010}. These publications provide the numerical framework required for the predictive simulation of plasma in a toroidal geometry. Within this category of modeling integration tools, TRANSP has become a full-featured system used for both interpretive and predictive analyses of tokamak plasmas~\cite{APankin2024}.

A significant issue for predictive simulation of transport processes is that physics processes create a stiffness that affects the numerical stability of transport solvers. In many applications, the nonlinear dependence of thermal and particle diffusivities, as well as the convective term, on local gradient values leads to potential difficulties with convergence, noise, and unintended numerical instability of simple implicit discretizations. Using an explicit approach, it has been demonstrated that gradient-dependent diffusion coefficients can produce severe numerical pathologies in one-dimensional diffusion situations~\cite{jardin08}. Another approach to address the numerical stability has been proposed by Pereverzev and Corrigan~\cite{Pereverzev2008}. The approach that they developed allows for significantly larger time steps compared to standard treatment and avoids artificial oscillations. These studies highlight the increasingly important role that numerical robustness and stability play in predictive transport modeling by demonstrating that they are not simply details of the implementation of the model, but actually represent a key element of the entire modeling effort.

To solve stiff transport equations, there are multiple possible algorithmic approaches. An efficient solver using the fourth-order interpolated differential operator (IDO) discretization with a method for nonlinear iteration that derives from a root-finding procedure has been recently developed~\cite{Park2017}. Using this method, both a plasma transport variable and its gradient are advanced, yielding more spatial accuracy on relatively coarse radial grids. Recently, Ludvig-Osipov {\em et al.} developed formulations for the transport equations in conservative form using fourth-order IDO discretization combined with second-order implicit Runge-Kutta time-stepping and under-relaxed Picard iteration~\cite{LudvigOsipov2025}. They focus on conservative properties, accuracy in transient periods, and robust convergence of solutions to stiff nonlinear transport models. These developments are examples of solver designs that employ higher-order discretizations combined with carefully chosen nonlinear iteration techniques to enhance accuracy and stability, while minimizing computation costs.

A somewhat different approach has been used for rapid predictive simulations for control, scenario optimization, and repeated model evaluation with the RAPTOR code.  Originally developed as a simulator for control, RAPTOR~\cite{Felici2018} generalized its application by providing the ability to simultaneously evolve the electron and ion densities and temperatures by using so called ``first-principles-based surrogate codes'' for rapid predictive simulations. Most of these applications have involved quick predictions of the performance of a set of parameters. Another enhancement to the code came with the implementation of the QLKNN model~\cite{VanMulders2021} for rapid optimization of stationary tokamak plasmas. In the case of the development of RAPTOR for tokamak operation using this new modeling system, the major objective was not to produce a fully integrated framework that utilizes both source and event physics, but rather to produce a workflow capable of efficiently making predictive assessments of data or models and assisting researchers to produce solutions more quickly than traditional approaches.

For higher levels of fidelity in physics, transport solvers have been directly coupled to gyrokinetic calculations. TGYRO~\cite{Candy2009} was developed for predicting steady-state profiles through direct gyrokinetic and neoclassical simulations. Another transport solver, TRINITY, supports time-dependent evolution of plasma profiles with turbulent fluxes computed with local nonlinear gyrokinetic turbulence simulations~\cite{Barnes2010}. These efforts have provided important benchmarks for modern predictive workflows and have drawn a useful distinction between steady-state and time-dependent transport solvers. They have also demonstrated how coupling to advanced transport models creates additional constraints on solver stability, nonlinear convergence, and the management of computational costs.

The goal of PT\_SOLVER is not just a high-fidelity gyrokinetic stand-alone transport solution, nor a real-time control application. Rather, PT\_SOLVER is intended to be an embedded, practical, robust, and implicit multi-channel transport solver within the integrated TRANSP environment, where the profile predictions are consistent with the models for sources, equilibrium computations, and current drive physics. Within the TRANSP framework, PT\_SOLVER has a clear path for validation against experiments due to the availability of the interpretive analysis within the same framework and multiple synthetic diagnostics, which make direct comparison with experiments straightforward. This objective sets the PT\_SOLVER requirements as robustness, modularity, and integration with other physics models in TRANSP.

As described in 2025 reference paper~\cite{APankin2024}, PT\_SOLVER supports several anomalous transport models, including GLF23~\cite{GLF23_1,GLF23_2}, TGLF~\cite{Kinsey2008TGLF,TGLF}, MMM~\cite{Rafiq:pop20,Rafiq:pop23}, RLW (Rebut-Lallia-Watkins model)~\cite{rebut89}, COPPI (Coppi-Tang model)~\cite{jardin93}, CDBM (Current Diffusive Ballooning Mode Model)~\cite{fukuyama95,takei07}, USER, and Paleo (Paleoclassical)~\cite{callen05} models, together with neoclassical models such as the modified Chang-Hinton model~\cite{chang86}, NEO~\cite{belli08,belli12}, and NCLASS~\cite{houlber97}. From the time of the TRANSP reference publication, TGLF has been extended to use saturation rule3 (SAT3)~\cite{Dudding_2022} and surrogates for TGLF (TGLF-NN)~\cite{Neiser2022_TGLFNNDatabase} and GKNN~\cite{Neiser2023_MultifidelityNN} have been implemented in PT\_SOLVER. Among anomalous models available in TRANSP, TGLF is particularly important in the present work because it provides a reduced, theory-based model of gyrokinetic turbulent transport while remaining practical enough for iterative predictive calculations. Its computational cost and evaluation over multiple flux surfaces motivate the multilevel parallelization strategy described below.

This paper describes the numerical scheme used for TRANSP predictive modeling. It does not attempt to present a comprehensive overview of all aspects of predictive workflows. Instead, it focuses on defining the PT\_SOLVER equations and the implicit Newton-based solution strategy to solve them; describing how coupled transport channels, source terms and stabilization methods are treated in TRANSP; and defining criteria for convergence in predictive simulations.  The verification of the solver is an important aspect of this work. The code-to-code comparisons using a common anomalous transport model, such as TGLF model~\cite{Staebler2005TGLF}, are especially valuable. They provide a direct test of solver behavior rather than comparisons in which both the transport physics and the numerical framework differ simultaneously. This approach is also consistent with recent cross-verification and validation studies emphasizing the need to distinguish numerical agreement from broader model-validation questions \cite{Abbate2024}.

In summary, there are three generic types of predictive transport approaches. The first one can be characterized as dynamic integrated modeling from a wider tokamak system perspective \cite{Jardin1986,Jardin2010}. The second group of applications focuses on developing a robust numerical method for time-stable solutions to stiff, nonlinear transport equations using unique discretization and iterative approaches for these equations \cite{jardin08,Pereverzev2008,Park2017,LudvigOsipov2025}. The third group focuses on either real-time, rapid control-oriented predictive simulations \cite{Felici2018,VanMulders2021}, or directly coupling algorithms to gyrokinetic turbulence simulations \cite{Candy2009,Barnes2010}.  PT\_SOLVER belongs most naturally to the first two classes. Its distinguishing role is to provide a robust, implicit, multi-channel transport capability inside TRANSP, with sufficient numerical stability to support predictive calculations using stiff transport closures in a mature integrated modeling environment.

\section{Governing Transport Equations and Assumptions}
The formulation assumes quasi-neutrality, with impurities in local thermodynamic equilibrium (LTE) using prescribed fractional abundances. Charge neutrality and effective charge ($Z_{\mathrm{eff}}$) constraints relate electron, ion, and impurity densities. The evolution of flux surfaces (via $\partial \xi/\partial t$) is provided by an equilibrium solver, and geometric metric factors are retained in the discretization. 

In a flux-surface -- averaged coordinate $\xi$ (normalized so that $\xi \in [0,1]$), the general conservation law for a plasma variable $U(\xi,t)$ is
\begin{equation}
  \frac{\partial U}{\partial t} + \frac{1}{V'}\frac{\partial}{\partial \xi}\Bigl\{ V'\,\mathfrak{F}(U) \Bigr\} = S(U),
  \label{eq:general}
\end{equation}
where $V'(\xi)=\frac{dV}{d\xi}$ is the differential volume (or area) element, $\mathfrak{F}(U)$ represents the total flux (including diffusive, convective, and grid-motion contributions), and $S(U)$ includes all source and sink terms.

The following equations give the schematic channel structure; the discrete implementation retains the full TRANSP flux-surface geometry, density weighting, source coupling, and moving-grid terms described in Sec.~\ref{sec:ptsolver}. In particular, the particle density $n(\xi,t)$ satisfies
\begin{equation}
  \frac{\partial n}{\partial t} + \frac{1}{V'} \frac{\partial}{\partial \xi}\left\{ V'\left[-D_n\,\frac{\partial n}{\partial \xi} + V_n\,n \right] - n\,\frac{\partial \xi}{\partial t} \right\} = S_n,
  \label{eq:density_cont}
\end{equation}
where $D_n$ is the diffusivity, $V_n$ is the convective (pinch) velocity, and $S_n$ describes sources such as recycling, neutral beam injection, and fusion reactions. PT\_SOLVER has separate options to advance densities for electrons, hydrogenic ions, and impurities. 


For electrons, the energy conservation equation is
\begin{equation}
  \frac{\partial (n_eT_e)}{\partial t} + \frac{1}{V'} \frac{\partial}{\partial \xi}\left\{ V'\left[-{\ap{n_e}} \chi_e\,\frac{\partial T_e}{\partial \xi} + V_{T_e}\,n_eT_e \right] - n_eT_e\,\frac{\partial \xi}{\partial t} \right\} = S_{T_e},
  \label{eq:electron_energy}
\end{equation}
and for ions,
\begin{equation}
  \frac{\partial (n_iT_i)}{\partial t} + \frac{1}{V'} \frac{\partial}{\partial \xi}\left\{ V'\left[-{\ap{n_i}} \chi_i\,\frac{\partial T_i}{\partial \xi} + V_{T_i}\,n_iT_i \right] - n_iT_i\,\frac{\partial \xi}{\partial t} \right\} = S_{T_i}.
  \label{eq:ion_energy}
\end{equation}
Here, $\chi_{e,i}$ denotes the effective thermal {\ap{diffusivities}} and $V_{T_{e,i}}$ the convective velocities. The source terms include collisional energy exchange: $Q_{ei} = n_e\,\nu_{ei}\,(T_e-T_i)$, where $\nu_{ei}$ is the electron-ion collision frequency; recycling contribution $Pe_{rcy}$ describing energy deposited by recycled neutrals; alpha particle heating $Pe_{\alpha}$; and other external sources, such as RF, neutral beam heating and radiative losses.

The toroidal angular momentum channel may be written schematically as
\begin{equation}
\frac{\partial p_\phi}{\partial t} +
\frac{1}{V'} \frac{\partial}{\partial \xi}
\left\{ V' \left[ -{\ap{\mathcal{I}}}_\phi \chi_\phi \frac{\partial \Omega_\phi}{\partial \xi} + V_\phi p_\phi
\right] - p_\phi \frac{\partial \xi}{\partial t} \right\} = S_\phi ,
\label{eq:pphi_transport}
\end{equation}

where {\ap{$p_\phi = \mathcal{I}_\phi \Omega_\phi$ is the flux-surface-averaged toroidal angular momentum density, $\Omega_\phi$ is the toroidal angular velocity, $\mathcal{I}_\phi = \sum_s m_s n_s \langle R^2\rangle$ is the moment-of-inertia density}}, $\chi_\phi$ is the momentum diffusivity, $V_\phi$ is the convective momentum pinch, and $S_\phi$ represents external torques and momentum sinks.

\section{PT\_SOLVER\label{sec:ptsolver}}
This section outlines the PT\_SOLVER numerical algorithm that is used to solve the transport equations that are described in the previous section. While the continuum formulation represents the essential physical content of the predictive problem, actual TRANSP simulations are dependent on the specific discretization that is contained within the code, including the implicit time-stepping scheme, linearization of transport operators, management of coupled sources and exchange terms, moving geometry effects, as well as the required nonlinear stabilization that is necessary for gradient-dependent transport coefficients. These terms are all significant because the PT\_SOLVER algorithm is not simply another implementation of a previously developed algorithm for gradient-dependent diffusion problems. The PT\_SOLVER algorithm is a code-specific implementation that is intended to solve robust multi-channel predictive problems within the TRANSP framework. As such, this section is intended to outline the discretized equations that are solved by PT\_SOLVER as well as the resultant convergence criteria that are used to solve the nonlinear problem.

In addition to the mathematical description below, PT\_SOLVER possesses certain other aspects of its design that are important from an engineering point of view. PT\_SOLVER is a modular, parallel, multi-regional transport solver that communicates with the rest of TRANSP through the Plasma State interface. It accommodates three radial regions with possibly different models being used in each region: axial region, confinement region, and edge region. The usefulness of this approach lies not only in making coding easier but also in allowing predictive calculations to stay valid even if a particular model is relevant only to one part of the plasma. Further, it offers a controlled way to include different models for different radial positions in a global calculation. This capability becomes especially important when considering situations in which the turbulence model underpredicts transport coefficients in the core or physics of the pedestal and near-axis requires different assumptions than those suitable for the confinement region. The boundary conditions when the multi-region approach is enabled are described in Table~\ref{tab:pt_regions_bc}. 

\begin{table}[!th]
\centering
\caption{Summary of PT\_SOLVER region definitions and boundary treatment in predictive TRANSP simulations. The notation $X_{\Upsilon,\mathrm{bound}}$ denotes the outer predictive boundary for channel $\Upsilon$, with $X_{T,\mathrm{bound}} \equiv XIBOUND$ for temperature channels, $X_{n,\mathrm{bound}} \equiv XNBOUND$ for density channels, and $X_{\phi,\mathrm{bound}} \equiv XPHIBOUND$ for toroidal rotation.}
\label{tab:pt_regions_bc}
\begin{tabular}{p{2.4cm} p{3.0cm} p{5.7cm} p{4.1cm}}
\hline
Element & Radial extent & Role in predictive calculation & Boundary treatment / typical control \\
\hline
Inner boundary & $\xi = 0$ & Magnetic-axis boundary for all evolved channels & Zero-gradient (Neumann-type) condition for evolved profiles \\
Axial region & $0 \le \xi < XIMIN\_CONF$ & Optional region used when alternate transport assumptions are needed near the magnetic axis, e.g.,\ where a confinement-region model is not reliable & Shares the inner boundary at $\xi=0$ and couples continuously to the confinement region at $\xi=XIMIN\_CONF$ \\
Confinement region & $XIMIN\_CONF \le \xi \le X_{\Upsilon,\mathrm{bound}}$ & Main predictive domain in which theory-based anomalous and neoclassical transport models are typically applied & Outer boundary for each channel is imposed at $X_{\Upsilon,\mathrm{bound}}$ using prescribed edge data, fixed values, or pedestal-model output \\
Edge region & $X_{\Upsilon,\mathrm{bound}} < \xi \le 1$ & Optional region used to represent the near-edge / pedestal domain with separate transport assumptions or prescribed profiles & Provides matching between the predictive solution and imposed edge constraints \\
\hline
\end{tabular}
\end{table}

PT\_SOLVER has a two-level parallel capability for selected models. In particular, the TGLF model implementation in PT\_SOLVER has one level of parallel processing for the transport solve for the individual flux surfaces, whereas another level can be used by TGLF, which has its own parallelism based on wavenumber decomposition. This approach utilizes sub-communicators for transport model calculations. After the TGLF computations are done, the information is combined by the solver for the Newton iteration process. There is more flexibility in terms of the number of processors available compared to approaches based only on the number of transport zones.

The main numerical challenge faced in predictive transport arises from the stiffness inherent to turbulence closures. Even in cases where the transport equation is not solved in terms of a threshold-based approach, there can be stiff behavior due to the strong dependence of the diffusivities and pinches on evolving local gradients. In addressing this problem, PT\_SOLVER uses a time-stepping algorithm based on the Newton method that is augmented with nonlinear stability conditions involving modified P\'{e}clet numbers, and with convergence conditions that consider both residuals and profile changes. The simpler problems of gradient-dependent diffusivities used in this paper do not constitute a replacement for the full TGLF framework, but serve merely as examples illustrating how stiffness can arise from transport parameters that depend strongly on evolving profiles.

\subsection{Numerical Algorithm\label{sec:ptnumerics}}
PT\_SOLVER uses an implicit time-advance combined with a Newton-based iterative method~\cite{jardin08}.

Assume the domain in $\xi$ is discretized at grid points $\xi_j$ with spacing $\Delta \xi$. A generic transport equation for a plasma variable is discretized via the following equation:
\begin{equation}
  \frac{U_j^{n+1} - U_j^n}{\Delta t} + \mathfrak{L}_j^{n+1}  = S_j,
  \label{eq:scheme}
\end{equation}
with the time-discrete algebraic transport operator at time $n+1$ approximated by
\begin{equation}
L^{n+1}_j = \frac{1}{\Delta t} \left(A_j U^{n+1}_{j-1} + B_j U^{n+1}_{j} + C_j U^{n+1}_{j+1}\right),
  \label{eq:spatial_operator}
\end{equation}
where $A_j$, $B_j$, and $C_j$ are dimensionless coefficients after the finite-volume and time-step scaling.

Separating implicit terms on the left-hand side and explicit terms on the right-hand side yields
\begin{equation}
U_j^{n+1}  + \mathfrak{L}_j^{n+1}{\Delta t} = - D_j \text{\ \ with } 
D_j = - U_j^n - S_j {\Delta t} .
\end{equation}
Here, \(D_j\) includes a combination of known terms at the previous timestep \( n \) and the explicitly known source terms \( S_j \). 
The coefficients $A_j$, $B_j$, $C_j$, and $D_j$ are derived from the discretization of the diffusive and convective fluxes and depend on local transport coefficients and geometric factors. 
For the electron temperature equation, they take on the following form:
\begin{align}
A_j &= -\frac{\Delta t}{1.5\, n_{e,j}\, V_j}\left[
\hat{\chi}_{e,j-\frac{1}{2}}\frac{S_{j-\frac{1}{2}}\,n_{e,j-\frac{1}{2}}}{\Delta^r_{j-\frac{1}{2}}}F({\ap{\mathbb{\hat P}}}_{e,j-\frac{1}{2}})\right]
+ \frac{\Delta t\, \dot{\xi}}{2\, n_{e,j}\, V_j}\,\xi_{j}\,n_{e,j}V'_{j}\label{eq:a},\\[12pt]
B_j &= 1 
+ \frac{\Delta t\, q_{ie,j}}{1.5\, n_{e,j}\, V_j\, k_B} + \frac{V_j - V_j^{(n)}}{1.5\, V_j}
+ \frac{\Delta t}{1.5\, n_{e,j}\, V_j}\left[
\hat{\chi}_{e,j+\frac{1}{2}}\frac{S_{j+\frac{1}{2}}\,n_{e,j+\frac{1}{2}}}{\Delta^r_{j+\frac{1}{2}}}
\left({\ap{\mathbb{\hat P}}}_{e,j+\frac{1}{2}}+F({\ap{\mathbb{\hat P}}}_{e,j+\frac{1}{2}})\right)\right.\nonumber\\
&\quad\left.
+ \hat{\chi}_{e,j-\frac{1}{2}}\frac{S_{j-\frac{1}{2}}\,n_{e,j-\frac{1}{2}}}{\Delta^r_{j-\frac{1}{2}}}F({\ap{\mathbb{\hat P}}}_{e,j-\frac{1}{2}})
\right]
- \frac{\Delta t\, \dot{\xi}}{2\, n_{e,j}\, V_j}\left[
\xi_{j+1} n_{e,j+1} V'_{j+1}
- \xi_{j} n_{e,j} V'_{j}
\right],\label{eq:b}\\[12pt]
C_j &= -\frac{\Delta t}{1.5\, n_{e,j}\, V_j}\left[
\hat{\chi}_{e,j+\frac{1}{2}}\frac{S_{j+\frac{1}{2}}\,n_{e,j+\frac{1}{2}}}{\Delta^r_{j+\frac{1}{2}}}F({\ap{\mathbb{\hat P}}}_{e,j+\frac{1}{2}})\right]
- \frac{\Delta t\, \dot{\xi}}{2\, n_{e,j}\, V_j}\,\xi_{j+1}\,n_{e,j+1}V'_{j+1},\label{eq:c}\\[12pt]
D_j &= \frac{\Delta t\,Q_{e,j}} {1.5\,n_{e,j} k_B V_j}  + \frac{\Delta t\,\dot{\xi}} {2 V_j} n_{e,j}T_{e,j}
\left(\xi_{j+1}V'_{j+1}-\xi_jV'_j\right)+ \frac{n_{e,j}T_{e,j}V_j}{\hat n_{e,j}\hat V_j},\label{eq:d}
\end{align}
where \(T_j\) and  \(n_{e,j}\) are the electron temperature and electron density at $j$ zone boundary at the previous time step,  \(n_{e,j}\) is the electron density at the current time step, \(\xi_j\) is the radial grid coordinate at $j$ zone boundary, \(\chi_{e,j+1/2}\) is the electron thermal diffusivity for $j$ zone  center,  \(S_{j\pm1/2}\) is the surface area at $j$ zone center, $V_j$ is the zone volume, $V'_j=\left.\left(\partial V/\partial \xi\right)\right|_j$ is the radial derivative of volume at $j$ zone boundary,  ${\ap{\mathbb{\hat P}}}_{e,j}$ is the modified electron P\'{e}clet number (which is discussed below), $q_{ie,j}$ is the electron-ion energy exchange rate at $j$ zone boundary, \(Q_{e,j}\) is the electron heating power density at spatial location \(j\), \(k_B\) is the Boltzmann constant, \(\dot{\xi}=\partial \xi/\partial t\) is term responsible for the grid radial motion, and $\Delta^r$ is the grid spacing defined as $d\xi/\bigl<|\nabla\xi|\bigr>$. 
In the coupled \(T_e/T_i\) solve, the electron-ion exchange term is treated implicitly through the block matrix. Thus, the term proportional to $q_{ie,j}(T_{e,j}-T_{i,j})$ contributes to the electron and ion diagonal
blocks and to the corresponding off-diagonal electron--ion coupling blocks.

The coefficients in Eqs.~\ref{eq:a}-\ref{eq:c} contain the stabilized conductivity
\begin{equation}
  \hat{\chi}_{e,j+1/2}  = \chi_{e,j+1/2} + \chi^{\rm stab}_{e,j+1/2},
\end{equation}
where \(\chi_{e,j+1/2}\) is the physical electron thermal diffusivity provided by the selected transport model, while \(\chi^{\rm stab}_{e,j+1/2}\) is a numerical stabilization parameter. In the present implementation,
\begin{equation}
  \chi^{\rm stab}_{e,j+1/2} =
  \max\!\left(f_{\chi,e}\,\chi^{\rm ref}_{e,j+1/2},
  \chi^{\min}_{e}\right),
\end{equation}
where $f_{\chi,e}$ and $\chi^{\min}_e$ are user-controlled numerical parameters, and $\chi^{\rm ref}_{e,j+1/2}$ is the local reference diffusivity used to scale the stabilization. The quantity \(\chi^{\rm stab}_{e}\) is not an additional physical transport coefficient. Rather, it enters the finite-difference operator to regularize the nonlinear solve in regions where the transport coefficients or profile gradients vary rapidly. 

{\ap{The purpose of $\chi_{e,j+1/2}^{\rm stab}$ is not to provide a small perturbation to the physical transport coefficient, but to define a numerical regularization scale used in the modified-P\'eclet discretization. It is part of the nonlinear discrete operator used to improve robustness of the implicit solve for gradient-sensitive transport models. The default values of $f_{\chi,e}$ and $\chi^{\rm min}_{e}$ are 1 and 0.01, respectively, a robustness-oriented numerical choice comparable to the diffusivities computed with a transport model. These default numerical values, $f_\chi=1$ and $\chi^{\min}=0.01$, are used for all transport channels unless channel-specific values are explicitly supplied by the user.  The stabilization is considered acceptable when the converged profiles, integrated fluxes, and power-balance residuals are insensitive to these changes within the accuracy required for the benchmark, while the nonlinear iteration remains robust. The default values of $f_{\chi,e}$ and $\chi^{\rm min}_{e}$ are selected based on these considerations. }}


At the domain boundaries (\( j = 1 \) and \( j = N \)), the coefficients of the tridiagonal system are modified to impose boundary conditions on the electron temperature. At the inner boundary (\( j = 1 \)), a zero-gradient (Neumann) condition is imposed:
\[
T_{e,2} = T_{e,1},
\]
which is implemented in the matrix system as:
\[
A_1 = -1, \quad B_1 = 1, \quad C_1 = 0, \quad D_1 = 0,
\]
enforcing \( -T_{e,1} + T_{e,2} = 0 \). At the outer boundary (\( j = N \)), a Dirichlet condition is applied:
\[
A_N = 0, \quad B_N = 1, \quad C_N = 0, \quad D_N = T_{e,\text{bdr}},
\]
where \( T_{e,\text{bdr}} \) is a specified boundary temperature at the plasma edge.

\begin{figure}[htbp]
  \centering
  \begin{tikzpicture}
    \begin{axis}[
      width=0.7\textwidth,
      height=0.5\textwidth,
      xlabel={$x = {\ap{\mathbb{\hat P}}}_e$},
      ylabel={$F(x)$},
      xmin=-15, xmax=15,
      ymin=-1, ymax=16,
      grid=both,
      domain=-15:15,
      samples=200,
      legend pos=south east,
      ]
      \addplot+[mark=none,blue] 
      { (x < -10 ? -x : ((x>=-10 && x<0) ? ((1+0.1*x)^5 - x) : ((x>=0 && x<10) ? ((1-0.1*x)^5) : 0)) ) };
    \end{axis}
  \end{tikzpicture}
  \caption{Plot of the stabilization function $F(x)$ versus $x={\ap{\mathbb{\hat P}}}_e$. For $x\ge10$, $F(x)=0$; for $x$ near 0, $F(x)\approx1$; and for $x<-10$, $F(x)=-x$.}
  \label{fig:stabilization}
\end{figure}
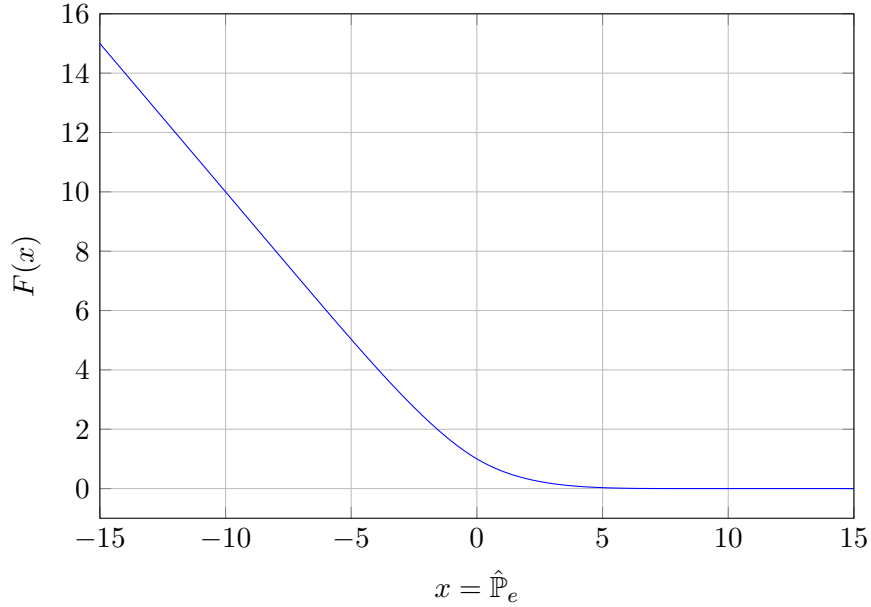

The same stabilization parameter also enters the modified P\'{e}clet number
used to control the upwind/central weighting of the face-centered
transport operator:
\begin{equation}
  {\ap{\mathbb{\hat P}}}_{e,j+1/2}
  =
  \frac{
  \left[
  V_{{\rm conv},j+1/2}
  +
  \chi^{\rm stab}_{e,j+1/2}
  \left(T'_e/T_e\right)_{j+1/2}
  \right]\Delta^r_{j+1/2}
  }
  {
  \chi_{e,j+1/2}+\chi^{\rm stab}_{e,j+1/2}
  } 
\end{equation}
This expression differs from the conventional P\'{e}clet number
\begin{equation}
  {\ap{\mathbb{P}}}_{e,j+1/2} = \frac{V_{{\rm conv},j+1/2}\Delta^r_{j+1/2}} {\chi_{e,j+1/2}},
\end{equation}
because it includes the additional numerical scale \(\chi^{\rm stab}_{e}\). 
This form makes the stabilization sensitive to the normalized profile gradient when the transport coefficients are small, rapidly varying, or strongly gradient-dependent. The modified P\'{e}clet number is then passed to the nonlinear stabilization function \(F({\ap{\mathbb{\hat P}}}_e)\), which appears directly in the discrete coefficients of the implicit solve:
\begin{equation}
  F(x) =
  \begin{cases}
    -x, & x < -10,\\[1mm]
    \left(1+0.1\,x\right)^5 - x, & -10 \le x < 0,\\[1mm]
    \left(1-0.1\,x\right)^5, & 0 \le x < 10,\\[1mm]
    0, & x \ge 10,
  \end{cases}
  \label{eq:stabilization_function}
\end{equation}
with $x = {\ap{\mathbb{\hat P}}}_e$. Figure~\ref{fig:stabilization} shows the plot of $F(x)$ versus $x={\ap{\mathbb{\hat P}}}_e$.  

The stabilization function varies nonlinearly with the modified P\'{e}clet number. In the limiting case of negligible physical convection and a stabilization scale that dominates the physical diffusivity, the modified P\'{e}clet number approaches a local normalized-gradient measure,
\[
{\ap{\mathbb{\hat P}}}_{e,j+1/2} \simeq
\left(\frac{T'_e}{T_e}\right)_{j+1/2}\Delta r_{j+1/2}.
\]
Thus, the stabilization acts most strongly in zones with large normalized gradients or rapidly varying transport coefficients. The numerical scale entering this expression is controlled by PT\_SOLVER input parameters and should be interpreted as a nonlinear stabilization device, not as an additional physical transport coefficient.

{\ap{Although the default value $f_{\chi,e}=1$ can make $\chi^{\rm stab}$ comparable to the physical diffusivity $\chi$, the effect of the stabilization on the converged solution is not determined by the ratio $\chi^{\rm stab}/\chi$ alone. The reason is that $\chi^{\rm stab}$ is not introduced as an independent physical diffusivity. It enters both the stabilized coefficient $\hat\chi=\chi+\chi^{\rm stab}$ and the modified P\'eclet number used in the face-centered discretization.}}

{\ap{This can be seen by considering a single face and neglecting physical convection for clarity. The face contribution to the discrete operator contains the combination
\begin{equation}
\hat\chi\left[{\ap{\mathbb{\hat P}}} T_L + F({\ap{\mathbb{\hat P}}})(T_L-T_R)\right],\label{eq:peclet-simple}
\end{equation}
where, for a smooth profile,
$$
{\ap{\mathbb{\hat P}}} \simeq \frac{\chi^{\rm stab}}{\hat\chi}\frac{T_R-T_L}{T_L}.
$$}}

{\ap{When the profile is resolved on the numerical grid, $|T_R-T_L|/T_L\ll 1$, so $|{\ap{\mathbb{\hat P}}}|\ll 1$ and $F({\ap{\mathbb{\hat P}}})\simeq 1$. In this limit, the term proportional to ${\ap{\mathbb{\hat P}}}$ cancels the leading artificial diffusive contribution associated with $\chi^{\rm stab}$, and the face operator reduces, to leading order, to the physical diffusive contribution proportional to $\chi$:
\begin{equation}
\hat\chi\left[{\ap{\mathbb{\hat P}}} T_L + F({\ap{\mathbb{\hat P}}})(T_L-T_R)\right]\simeq -\chi^{\rm stab} (T_L-T_R) + (\chi + \chi^{\rm stab}) (T_R-T_L) = \chi (T_R-T_L)
\end{equation}
Therefore, even when $\chi^{\rm stab}$ is comparable to $\chi$, the resulting correction to the numerical flux is controlled primarily by the cell-scale normalized gradient and by the departure of $F({\ap{\mathbb{\hat P}}})$ from its small ${\ap{\mathbb{\hat P}}}$ limit, not simply by $\chi^{\rm stab}/\chi$.}}

{\ap{The stabilization becomes active when the modified P\'eclet number is not small, which occurs in regions with steep gradients, small or rapidly varying physical transport coefficients, or insufficiently resolved local profile structure. In those regions the modified operator is intended to improve robustness of the nonlinear solve. For this reason, the acceptability of the stabilization is assessed by sensitivity tests in which $f_\chi$, $\chi^{\min}$, and the grid resolution are varied. The stabilization is considered not to affect the physical conclusions when the converged profiles, integrated fluxes, and power-balance residuals are insensitive to these changes within the accuracy required for the benchmark.}}

{\ap{These arguments about the effect of $\chi^{\rm stab}$ can help to understand our motivations in the selection of a particular form  for $F({\ap{\mathbb{\hat P}}})$.}}

{\ap{The particular form of $F({\ap{\mathbb{\hat P}}})$ is not intended to represent an additional physical closure. It is a numerical limiter used in the modified-P\'eclet discretization. Its form was chosen to satisfy several practical consistency and robustness requirements. First, $F(0)=1$, so that in the limit of a smooth resolved profile and small modified P\'eclet number, the face-centered operator reduces to the physical diffusive operator. In this limit, the leading artificial diffusive contribution associated with $\chi^{\rm stab}$ cancels between the ${\ap{\mathbb{\hat P}}}$-dependent part of the operator and the term proportional to $F({\ap{\mathbb{\hat P}}})(T_L-T_R)$. Thus, even when $\chi^{\rm stab}$ is comparable to $\chi$, the small $|{\ap{\mathbb{\hat P}}}|$ limit is consistent with the physical diffusion coefficient $\chi$.}}

{\ap{Second, the polynomial transition gives a smooth variation of the discrete coefficients as ${\ap{\mathbb{\hat P}}}$ changes sign or approaches the limiting values. This is important for Newton iteration because abrupt changes in the differencing formula would introduce additional nonsmoothness into an already nonlinear transport problem. Third, the limiting values $F({\ap{\mathbb{\hat P}}})=0$ for large positive ${\ap{\mathbb{\hat P}}}$ and $F({\ap{\mathbb{\hat P}}})=-{\ap{\mathbb{\hat P}}}$ for large negative ${\ap{\mathbb{\hat P}}}$ produce a more one-sided, upwind-like operator in regimes with steep gradients, rapidly varying transport coefficients, or large effective convective drive. These are precisely the regimes in which the central-like small $|{\ap{\mathbb{\hat P}}}|$ operator is least robust.}}

{\ap{The fifth-power form used here is therefore best regarded as a code-specific smooth limiter satisfying the required small ${\ap{\mathbb{\hat P}}}$ consistency, smooth-transition, and large $|{\ap{\mathbb{\hat P}}}|$ stabilization properties. It is not unique, and other limiter functions with the same limiting behavior could be constructed. For this reason, the physical acceptability of the stabilization is assessed not by the pointwise condition $\chi^{\rm stab}\ll\chi$, but by sensitivity tests in which the stabilization parameters and grid resolution are varied and the converged profiles, integrated fluxes, and residuals are compared.}}

The residual at each grid point is defined as

\begin{equation}
  \mathfrak{R}_j = \frac{U_j^{n+1} - U_j^n}{\Delta t} + \frac{1}{V'_j}\Bigl(\tilde{A}_j\,U_{j-1}^{n+1} + B_j\,U_j^{n+1} + \tilde{C}_j\,U_{j+1}^{n+1}\Bigr) - S_j.
  \label{eq:residual_newton}
\end{equation}
The L2 norm of the residual is then
\begin{equation}
  \| \mathbf{R} \|_2 = \left( \sum_{j=1}^{N} \mathfrak{R}_j^2\,\Delta \xi \right)^{1/2}.
  \label{eq:L2_norm_final}
\end{equation}

To monitor convergence in the electron energy solver, PT\_SOLVER evaluates the signed discrete energy-balance defect in each radial zone.  We first define
\begin{align}
  Q_{1,j} &= P_{e,j} - \frac{\partial (n_eT_e)_j}{\partial t} - Q_{ei,j} - Q_{\text{cmp},j} - Q_{\text{cmp,eldot},j} - Q_{\text{eldot},j}, \label{eq:Q1}\\[1mm]
  Q_{2,j} &= Q_{e,j}^{\mathrm{ext}} - Q_{e,j}^{\mathrm{in}}, \label{eq:Q2}
\end{align}
where {\ap{$Q_{1,j}$ and $Q_{2,j}$ are zone-integrated power flows,}} $P_{e,j}$ is the zone-integrated external power input to electrons, including contributions from neutral beam injection, recycling, and alpha-particle heating; \({\partial (n_eT_e)_j}/{\partial t}\) is the time derivative of the electron energy density in zone $j$ with $n_eT_e$ representing the electron energy density; $Q_{ei,j}$ is the collisional energy exchange between electrons and ions;  \(Q_{\text{cmp},j}\) 
represents the plasma compression due to plasma volume change:
\begin{equation}
Q_{\text{cmp},j} =\frac{k_B\, n_{e,j}\, T_{e,j}}{\Delta t} \left( V_j - V_j^{(n)} \right)
\end{equation}
\(Q_{\text{cmp,eldot},j}\) is the grid-motion correction to the compression term due to grid motion-induced volume change:
\begin{equation}
Q_{\text{cmp,eldot},j} =k_B \dot{\xi}  \left(\xi_{j+1} V'_{j+1}- \xi_j V'_j\right) n_{e,j} T_{e,j}
\end{equation}
\(Q_{\text{eldot},j}\) is the energy contribution from the net pressure work due to mesh velocity:
\begin{equation}
Q_{\text{eldot},j} =-\frac{3}{2} k_B \dot{\xi} \left( \xi_{j+1} V'_{j+1} \frac{ n_{e,j} T_{e,j} + n_{e,j+1} T_{e,j+1}}{2} - \xi_{j} V'_{j} \frac{ n_{e,j-1} T_{e,j-1} + n_{e,j} T_{e,j}}{2} \right)
\end{equation}
\(Q_{e,j}^{\mathrm{ext}}\) and \(Q_{e,j}^{\mathrm{in}}\) are the transport fluxes evaluated at the exterior and interior boundaries of zone \(j\), respectively. Their difference \(Q_{2,j}\) 
provides the net transport contribution across the zone.

The local electron energy-balance defect is then
\begin{equation}
  \epsilon_{e,j} = Q_{2,j}-Q_{1,j},
  \label{eq:electron_balance_defect}
\end{equation}
and the corresponding unsigned local power budget used for normalization is
\begin{align}
  B_{e,j} ={}& \left|Q_{e,j}^{\mathrm{ext}}\right| + \left|Q_{e,j}^{\mathrm{in}}\right| + \left|P_{e,j}\right|
  + \left|\frac{\partial (n_eT_e)_j}{\partial t}\right| + \left|Q_{ei,j}\right| \\ 
  &+ \left|Q_{\text{cmp},j}\right| + \left|Q_{\text{cmp,eldot},j}\right| + \left|Q_{\text{eldot},j}\right| .
  \label{eq:electron_balance_budget}
\end{align}
The normalized electron-temperature residual is defined as
\begin{equation}
\mathbb{R}_{T_e} =
\frac{\left(\sum_{j=1}^{N} \epsilon_{e,j}^{2}\right)^{1/2}}
{\max\left[\left(\sum_{j=1}^{N} B_{e,j}^{2}\right)^{1/2},\,B_{\rm floor}\right]} .
\label{eq:normalized_residual_te}
\end{equation}
A pointwise residual may also be monitored as
\begin{equation}
  r_{T_e,j}=\frac{|\epsilon_{e,j}|}{\max\left(B_{e,j},B_{\rm floor}\right)} .
  \label{eq:pointwise_residual_te}
\end{equation}
Here \(B_{\rm floor}\) is a small positive floor that prevents division by zero in zones where all terms in the local balance are negligible.  This formulation measures the normalized defect of the signed discrete conservation law.  The normalization is based on the unsigned local power budget rather than on the sum of two signed net quantities, avoiding artificial saturation when the source/time-derivative side and the transport-flux side have opposite signs.

The fluxes \(Q^{\rm ext}_{e,j}\) and \(Q^{\rm in}_{e,j}\) entering Eqs.~(\ref{eq:electron_balance_defect})--(\ref{eq:pointwise_residual_te}) are evaluated from the same face-centered discrete operator that is used to construct the implicit tridiagonal system. Thus, when modified-P\'{e}clet stabilization is enabled, the residual is formed using the stabilized numerical flux, not by recomputing the flux from only the unstabilized physical conductivity. Schematically, the numerical flux used in the nonlinear solve may be
written as
\[
  Q^{\rm num}_{e,j+1/2}
  = Q^{\rm phys}_{e,j+1/2} + Q^{\rm stab}_{e,j+1/2},
\]
where \(Q^{\rm phys}_{e,j+1/2}\) contains the physical conductive and convective terms, while \(Q^{\rm stab}_{e,j+1/2}\) denotes the contribution introduced by the modified-P\'{e}clet discretization through $\chi^{\rm stab}_e$ and $F({\ap{\mathbb{\hat P}}}_e)$. The residual used for nonlinear convergence is therefore the defect of the stabilized discrete conservation law,
\[
  \epsilon_{e,j}
  = \left(Q^{\rm num,ext}_{e,j}-Q^{\rm num,in}_{e,j}\right) - Q_{1,j}.
\]
This choice ensures that convergence is tested against the same algebraic system that is being solved. A residual reconstructed from \(Q^{\rm phys}_{e}\) alone is still useful as a diagnostic of the size of the stabilization correction and of the physical power-balance closure, but it is not the residual of the nonlinear discrete problem when stabilization is active.

The same consistency requirement applies to the moving-grid and compression terms: the face interpolation used in the residual reconstruction is kept identical to that used in the implicit coefficient assembly, so that the residual vanishes for a converged solution of the discretized equations up to solver tolerance.

Residuals are determined separately for each transport channel. Similarly, the profile change ratio $\mathfrak{C}_i$ is calculated for each channel that is predicted, with $\mathfrak{C}_{\max}$ being equal to $\max_i(\mathfrak{C}_i)$. Convergence of the nonlinear iteration is established if $\mathfrak{C}_{\max} < \mathfrak{C}_s$ and $\mathbb{R}_i < \mathfrak{r}_i$ for each transport channel $i$, where $\mathfrak{r}_i$ is the user-defined threshold value for residual for channel $i$, and $\mathfrak{C}_s$ is the threshold value for the maximum allowable profile change ratio.

For production runs, convergence depends on the simultaneous satisfaction of channel-specific residual tolerance criteria as well as the maximum profile change criteria. In order to increase robustness in the context of long-term simulation prediction runs, PT\_SOLVER contains fail-safe procedures for those cases in which the adaptive time stepping procedure approaches the minimum time step specified by the user. In such situations, the solver can take a time step that meets an explicitly stated emergency criterion, on condition that profile changes are bounded and residuals do not exceed channel-specific multiples of the standard tolerances. Such time steps are tagged in the output and should be distinguished from steps satisfying the standard nonlinear convergence criterion. They are not used to define the strict convergence behavior reported in the verification tests.

\subsection{Testing the Convergence of the Numerical Algorithm\label{sec:pttest}}

In order to test the numerical convergence of the PT\_SOLVER numerical scheme, we have developed a Python script that solves a simplified family of coupled one-dimensional nonlinear transport equations designed to reproduce the main numerical features of PT\_SOLVER, including multi-channel coupling, implicit time advancement, Newton iterations, and modified-P\'eclet stabilization. It is not intended to reproduce the full physical TRANSP formulation in flux-surface geometry, but rather to provide a flexible numerical testbed for studying stability, convergence, and coupling behavior. For reproducibility of the results, the Python script, its inputs, and outputs are uploaded to Princeton Data Commons and available for download using the URL shared at the end of this article in the Data Availability section.

The model equations can be written as

\begin{equation}
\frac{\partial U_m}{\partial t}+
\frac{\partial}{\partial x}
\left[
- D_m\!\left(U,\partial_x U\right)\frac{\partial U_m}{\partial x}
+ v_m\!\left(U,\partial_x U\right) U_m
- \dot{x}\,U_m
\right]
=
S_m^{(\mathrm{par})}(x,t)+\sum_{j\neq m} c_{mj}\,\bigl(U_j-U_m\bigr),
\end{equation}
where $m=1,\dots,M$.

Here, $U_m(x,t)$ is the $m$-th transport channel, $D_m$ is the channel diffusivity, $v_m$ is the convective velocity, $\dot{x}$ is an optional mesh velocity term, $Sm^{\rm(par)}$ is the prescribed parametrized source, and $c_{mj}$ are optional linear coupling coefficients between channels.

The two diffusivity models in the script are

\begin{equation}
D_m^{\mathrm{linear}} = D_{0,m} + a_m \left| \frac{\partial U_m}{\partial x} \right|,
\end{equation}

and

\begin{equation}
D_m^{\mathrm{stiff}}=D_{0,m}+a_m \max\!\left(\left| \frac{\partial U_m}{\partial x} \right|-g_{\mathrm{crit},m},0\right).
\end{equation}

With stabilization enabled, the script actually uses an effective diffusivity

\begin{equation}
D_{m,\mathrm{eff}}
=D_m + \chi^{\rm stab}_m\,F({\ap{\mathbb{\hat P}}}_m),
\end{equation}

with

\begin{equation}
{\ap{\mathbb{\hat P}}}_m =
\frac{\left(v_m + \chi^{\rm stab}_m\,\partial_x U_m / U_m\right)\Delta x}{
D_m + \chi^{\rm stab}_m},
\end{equation}
and the same piecewise stabilization function $F$ defined in Sec.~\ref{sec:ptnumerics}. The script is therefore aligned with the numerical idea of modified-P\'eclet stabilization, but in a simplified setting.

\subsubsection{Analytical benchmark: coupled diffusion with constant exchange}
\label{sec:analytic_benchmark}

To verify the numerical implementation, we consider a two-channel linear transport problem for
electron and ion temperatures, $T_e(x,t)$ and $T_i(x,t)$, on the interval
$0 \le x \le L$, with constant diffusivities and symmetric inter-channel relaxation:
\begin{align}
\frac{\partial T_e}{\partial t}
&=
D \frac{\partial^2 T_e}{\partial x^2}
+
\alpha \left(T_i-T_e\right),
\label{eq:Te_benchmark}
\\
\frac{\partial T_i}{\partial t}
&=
D \frac{\partial^2 T_i}{\partial x^2}
+
\alpha \left(T_e-T_i\right),
\label{eq:Ti_benchmark}
\end{align}
where $D$ is the constant diffusivity and $\alpha$ is the coupling coefficient.

Homogeneous Dirichlet boundary conditions are imposed at both ends of the domain,
\begin{equation*}
T_e(0,t)=T_e(L,t)=0,
\qquad
T_i(0,t)=T_i(L,t)=0,
\label{eq:benchmark_bc}
\end{equation*}
and the initial conditions are chosen as a single sinusoidal mode,
\begin{equation*}
T_e(x,0)=A_e \sin(kx),
\qquad
T_i(x,0)=A_i \sin(kx),
\qquad
k=\frac{n\pi}{L}.
\label{eq:benchmark_ic}
\end{equation*}

It is convenient to introduce the sum and difference variables
\begin{equation*}
S = T_e + T_i,
\qquad
\Delta = T_e - T_i.
\label{eq:sumdiff_def}
\end{equation*}
Substituting Eqs.~\eqref{eq:Te_benchmark}--\eqref{eq:Ti_benchmark} gives two decoupled equations,
\begin{equation*}
\frac{\partial S}{\partial t}=D \frac{\partial^2 S}{\partial x^2},
\qquad
\frac{\partial \Delta}{\partial t}=D \frac{\partial^2 \Delta}{\partial x^2}-2\alpha \Delta.
\end{equation*}

Since the initial condition contains only a single eigenmode, the exact solution remains
proportional to $\sin(kx)$ for all time:
\begin{equation*}
S(x,t)=(A_e+A_i)\,e^{-Dk^2 t}\sin(kx),
\qquad
\Delta(x,t)=(A_e-A_i)\,e^{-(Dk^2+2\alpha)t}\sin(kx).
\end{equation*}

The exact solutions for the two channels then follow from
$T_e=(S+\Delta)/2$ and $T_i=(S-\Delta)/2$:
\begin{equation*}
T_{e,i}(x,t)=
\frac{1}{2}\left[(A_{e,i}+A_{i,e})e^{-Dk^2 t}
\pm(A_{e,i}-A_{i,e})e^{-(Dk^2+2\alpha)t}\right]\sin(kx),
\label{eq:Te_exact}
\end{equation*}

\begin{figure}[h!]
  \centering
  \includegraphics[width=0.95\textwidth]{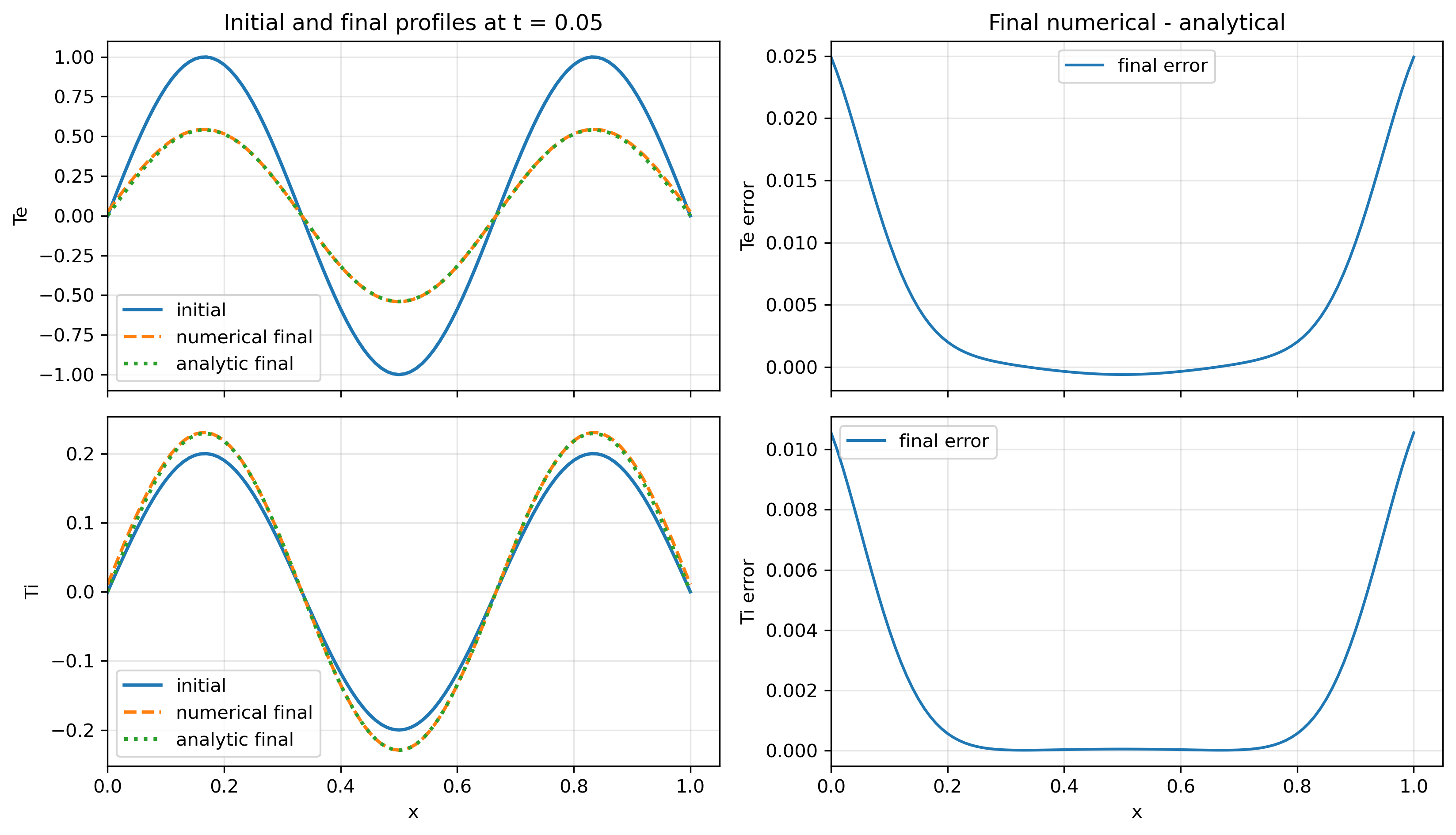}
  \caption{Initial and final profiles for the constant-diffusivity coupled benchmark with the following parameters: $L=1$, $n=3$, $D=0.10$, $\alpha=5.0$, $A_e=1.0$, $A_i=0.2$, and $k=3\pi$. Shown are the initial electron and ion temperature profiles and the corresponding numerical solutions at the final time. The profile evolution is governed by constant diffusion together with symmetric electron--ion relaxation, so that both overall decay and equilibration between channels are visible. The left panels show the initial, final, and analytical solutions for electron and ion temperature profiles, and the right panels show the absolute errors between the numerical and analytical solutions.}
  \label{fig:constant_diffusivity}
\end{figure}

For the specific parameters given in the caption to Fig.~\ref{fig:constant_diffusivity}, the solution becomes
\begin{equation*}
T_{e,i}(x,t)=\left[0.6\,e^{-0.9\pi^2 t}\pm0.4\,e^{-(0.9\pi^2+10)t}\right]\sin(3\pi x),
\label{eq:Te_exact_specific}
\end{equation*}
The results of numerical and analytical solution comparison are shown in Fig.~\ref{fig:constant_diffusivity}.
This benchmark is useful because it tests both diffusion and coupled relaxation in a setting
with a closed-form analytical solution. The common mode, $S=T_e+T_i$, decays purely by diffusion,
while the difference mode, $\Delta=T_e-T_i$, decays faster due to both diffusion and
inter-channel exchange.

\subsubsection{Manufactured-solution test with evolving H-mode pedestal}
\label{sec:mms_hmode_evolving}
The previous test demonstrates that the coupled diffusion equations evolved to the analytical solution. However, the problem was formulated using simple sinusoidal profiles with
homogeneous boundary conditions, which are far from any experiment-relevant observations. 

To verify the coupled transport solver for profile shapes more representative of tokamak applications, we consider a method-of-manufactured-solutions (MMS) test for electron and ion temperatures, $T_e(x,t)$ and $T_i(x,t)$, on the interval $0 \le x \le L$. The governing equations are
\begin{align}
\frac{\partial T_e}{\partial t}
&=D_e \frac{\partial^2 T_e}{\partial x^2}
+\alpha \left(T_i-T_e\right)
+S_e(x,t),
\label{eq:mms_evolving_te}
\\
\frac{\partial T_i}{\partial t}
&=D_i \frac{\partial^2 T_i}{\partial x^2}
+\alpha \left(T_e-T_i\right)
+S_i(x,t),
\label{eq:mms_evolving_ti}
\end{align}
where $D_e$ and $D_i$ are constant diffusivities and $\alpha$ is the inter-channel relaxation coefficient.

The exact manufactured solutions are chosen as H-mode-like profiles with a time-dependent pedestal amplitude and an additional decaying sinusoidal perturbation,
\begin{equation}
T_m^{\mathrm M}(x,t)
=
T_{m,\mathrm{ped}}
+
\Delta_m(t)\,H(x)
+
a_m e^{-\lambda_m t}\sin(k_m x),
\qquad
m\in\{e,i\},
\label{eq:mms_evolving_exact}
\end{equation}
with
\begin{equation}
H(x)
=
\frac{1}{2}
\left[
1-\tanh\!\left(\frac{x-x_{\mathrm{ped}}}{w}\right)
\right],
\label{eq:mms_evolving_H}
\end{equation}
\begin{equation*}
\Delta_m(t)
=\Delta_{m,0} + \Delta_{m,1}\left(1-e^{-\beta_m t}\right),
\label{eq:mms_evolving_delta}
\end{equation*}
and $k_m = {n_m \pi}/{L}.$

The pedestal shape is controlled by the fixed function $H(x)$, while its amplitude evolves in time through $\Delta_m(t)$. The sine term provides an additional smooth perturbation that decays exponentially in time.

The source terms are constructed analytically by substituting the manufactured solutions into Eqs.~\eqref{eq:mms_evolving_te}--\eqref{eq:mms_evolving_ti},
\begin{align}
S_e(x,t)
&=\frac{\partial T_e^{\mathrm M}}{\partial t}
-D_e \frac{\partial^2 T_e^{\mathrm M}}{\partial x^2}
-\alpha \left(T_i^{\mathrm M}-T_e^{\mathrm M}\right),
\label{eq:mms_evolving_Se_def}
\\
S_i(x,t)
&=\frac{\partial T_i^{\mathrm M}}{\partial t}
-D_i \frac{\partial^2 T_i^{\mathrm M}}{\partial x^2}
-\alpha \left(T_e^{\mathrm M}-T_i^{\mathrm M}\right).
\label{eq:mms_evolving_Si_def}
\end{align}

To evaluate these expressions, note that $\dot{\Delta}_m(t)=\Delta_{m,1}\beta_m e^{-\beta_m t}$,
and, with $\eta = ({x-x_{\mathrm{ped}}})/{w}$, the second derivative of the pedestal shape is
\begin{equation*}
H''(x)
=
\frac{1}{w^2}\,\mathrm{sech}^2(\eta)\tanh(\eta).
\label{eq:mms_evolving_Hpp}
\end{equation*}
Therefore,
\begin{align}
\frac{\partial T_m^{\mathrm M}}{\partial t}
&=
\dot{\Delta}_m(t)\,H(x)
-
\lambda_m a_m e^{-\lambda_m t}\sin(k_m x),
\\
\frac{\partial^2 T_m^{\mathrm M}}{\partial x^2}
&=
\Delta_m(t)\,H''(x)
-
a_m k_m^2 e^{-\lambda_m t}\sin(k_m x).
\label{eq:mms_evolving_derivs}
\end{align}

The benchmark is solved with exact Dirichlet boundary conditions,
\begin{equation*}
T_e(0,t)=T_e^{\mathrm M}(0,t),
\qquad
T_e(L,t)=T_e^{\mathrm M}(L,t),
\qquad
T_i(0,t)=T_i^{\mathrm M}(0,t),
\qquad
T_i(L,t)=T_i^{\mathrm M}(L,t),
\label{eq:mms_evolving_bc}
\end{equation*}
and with initial conditions taken directly from the manufactured solution at $t=0$,
\begin{equation*}
T_e(x,0)=T_e^{\mathrm M}(x,0),
\qquad
T_i(x,0)=T_i^{\mathrm M}(x,0).
\label{eq:mms_evolving_ic}
\end{equation*}

In the present test, the numerical values are
\begin{equation*}
L=1, \qquad x_{\mathrm{ped}}=0.82, \qquad w=0.04,
\end{equation*}
\begin{equation*}
D_e = 0.05, \qquad D_i = 0.03, \qquad \alpha = 3.0,
\end{equation*}
and
\begin{align*}
T_{e,\mathrm{ped}} &= 0.30, &
\Delta_{e,0} &= 0.55, &
\Delta_{e,1} &= 0.40, &
\beta_e &= 20.0, &
a_e &= 0.08, &
\lambda_e &= 6.0, &
n_e &= 2,
\\
T_{i,\mathrm{ped}} &= 0.18, &
\Delta_{i,0} &= 0.35, &
\Delta_{i,1} &= 0.28, &
\beta_i &= 14.0, &
a_i &= -0.06, &
\lambda_i &= 5.0, &
n_i &= 3.
\end{align*}

The simulation is performed on a uniform grid with
\begin{equation*}
N_x = 201, \qquad \Delta t = 10^{-4}, \qquad t_{\mathrm{final}} = 5\times10^{-2}.
\end{equation*}

\begin{figure}[htbp]
  \centering
  \includegraphics[width=0.95\textwidth]{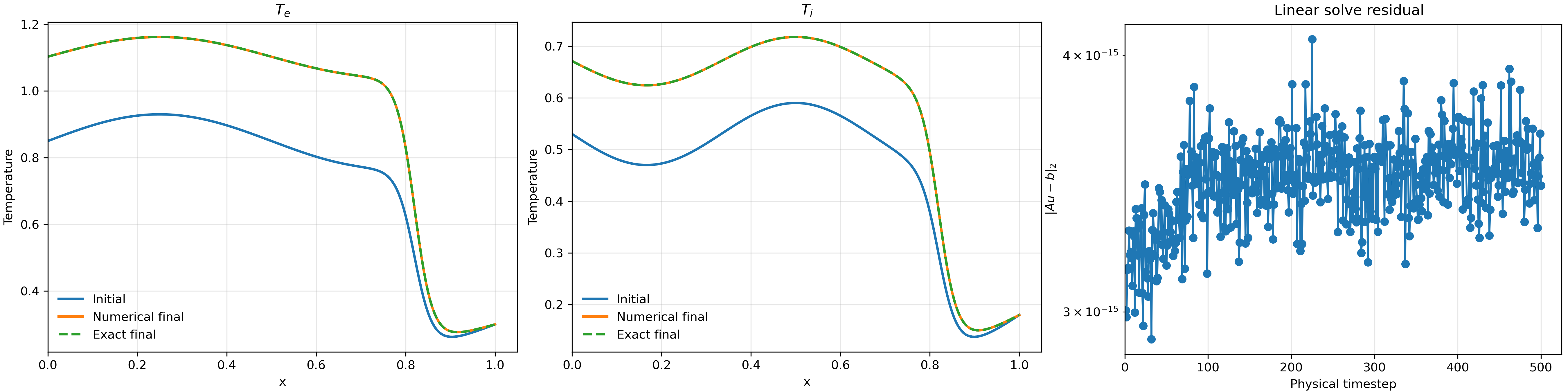}
\caption{MMS benchmark with evolving H-mode-like pedestal profiles. Left and middle panels compare the initial profiles with the numerical and exact manufactured solutions at the final time for the electron and ion temperature channels, respectively. The benchmark combines constant diffusivities, inter-channel relaxation, exact source balancing, and time-dependent pedestal amplitudes, so that the full profile shape evolves while remaining analytically prescribed. The numerical and exact final profiles are nearly indistinguishable on the scale of the plot, indicating accurate reproduction of the manufactured solution. The right panel shows the residual of the linear solve at each physical time step, demonstrating stable and well-converged implicit advancement throughout the simulation.}
  \label{fig:mms}
\end{figure}

The results of our test are summarized in Fig.~\ref{fig:mms}. This MMS test is more demanding than a single-mode diffusion benchmark because it combines nontrivial H-mode-like spatial structure, time-dependent pedestal evolution, inter-channel coupling, and analytically known source terms. It therefore provides a stringent verification case for both the spatial discretization and the time advancement of the coupled solver.

\subsubsection{Numerical test for the model with gradient-dependent diffusivities}
Physics-based models for anomalous transport incorporate the dependence of effective diffusivities on plasma gradients. Typically, the dependence is nonlinear. A particular case of effective diffusivity that depends linearly on plasma gradients and has a stability threshold can be described with this expression:
\begin{equation}
D_m^{\mathrm{stiff}}
= D_{0,m} + a_m \max\!\left(\left| \frac{\partial U_m}{\partial x} \right|
-g_{\mathrm{crit},m}, 0 \right).\label{eq:stiffmodel}
\end{equation}
In this expression, $D_{0,m}$ is the minimal diffusivity that is always present and independent of the plasma gradients ${\partial U_m}/{\partial x}$, $g_{\mathrm{crit},m}$ is the critical gradient level that represents the threshold of instability that drives the anomalous transport, and $a_m$ is the stiffness coefficient. 
This stiff transport model is representative of a wide class of drift-wave instabilities in tokamaks. There is no exact analytical solution for this case, and we present results of convergence studies for a particular selection of diffusivities in coupled plasma profile equations. 

\begin{wrapfigure}{r}{0.5\textwidth} 
\vspace{-20pt}
  \begin{center}
    \includegraphics[width=0.48\textwidth]{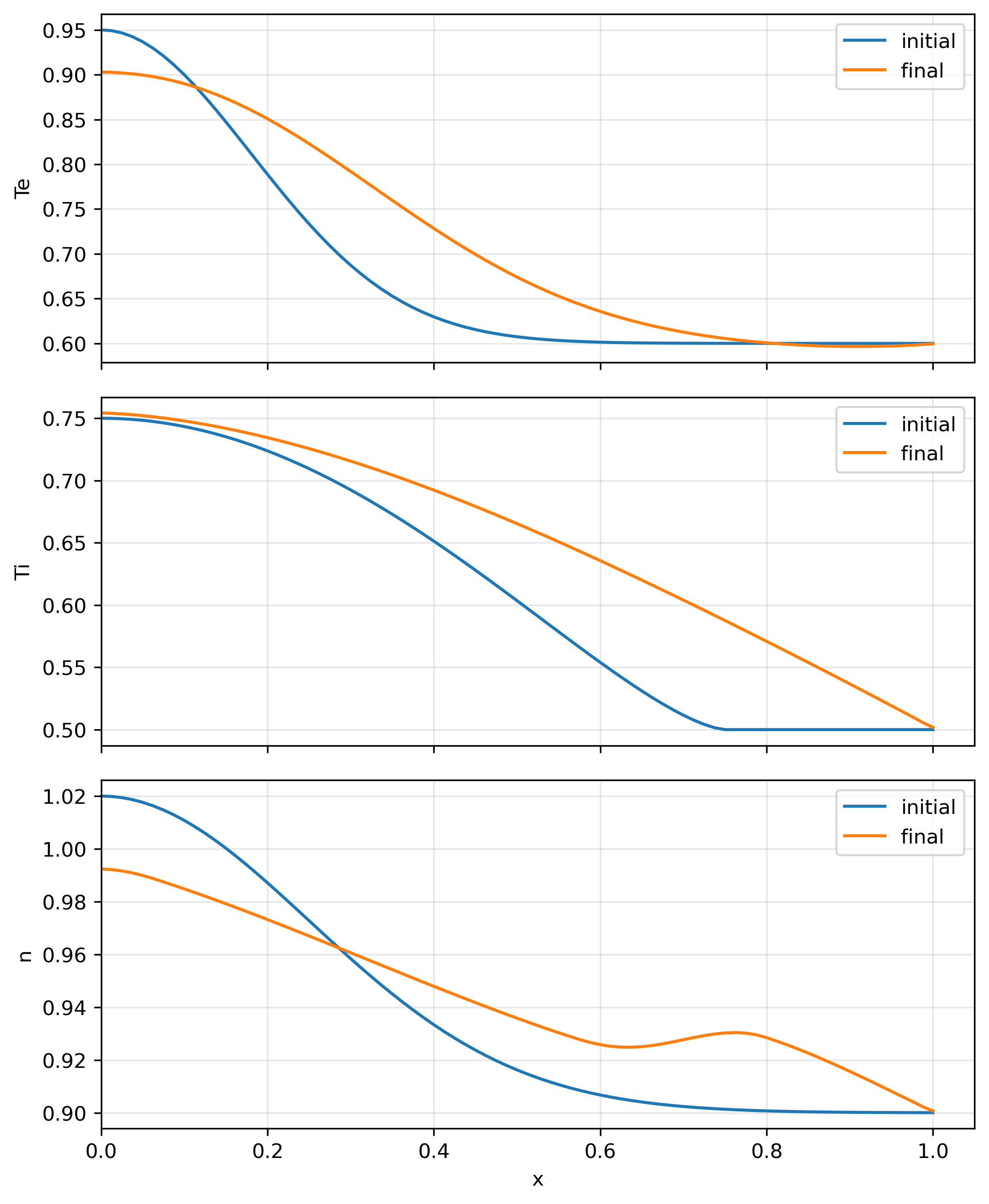}
    \vspace{-10pt}
    \caption{Profiles evolution for the coupled three-channel transport test at initial and final time steps. Results are shown for four physical time steps, $\Delta t=5\times10^{-4}$, $10^{-3}$, $2\times10^{-3}$, and $10^{-2}$. The results converge to the same final states independent of the physical time steps. The residuals for different iterations are shown in the left panels of Fig.~\ref{fig:dd_converge}. }
    \label{fig:dd_profiles}
    \vspace{-25pt}
  \end{center}
\end{wrapfigure}

\begin{figure}[htbp]
 \centering
  \includegraphics[width=0.45\textwidth]{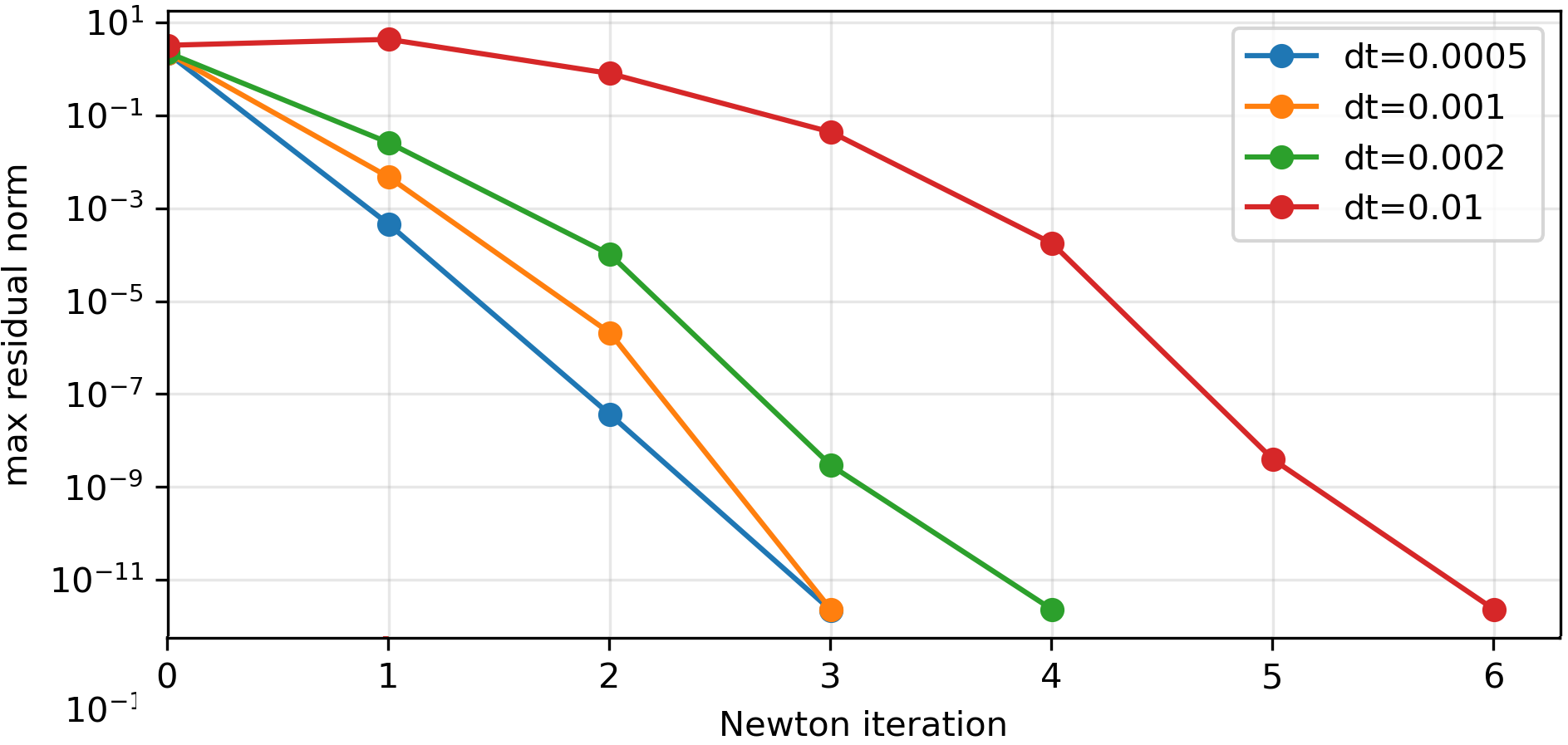}
  \includegraphics[width=0.45\textwidth]{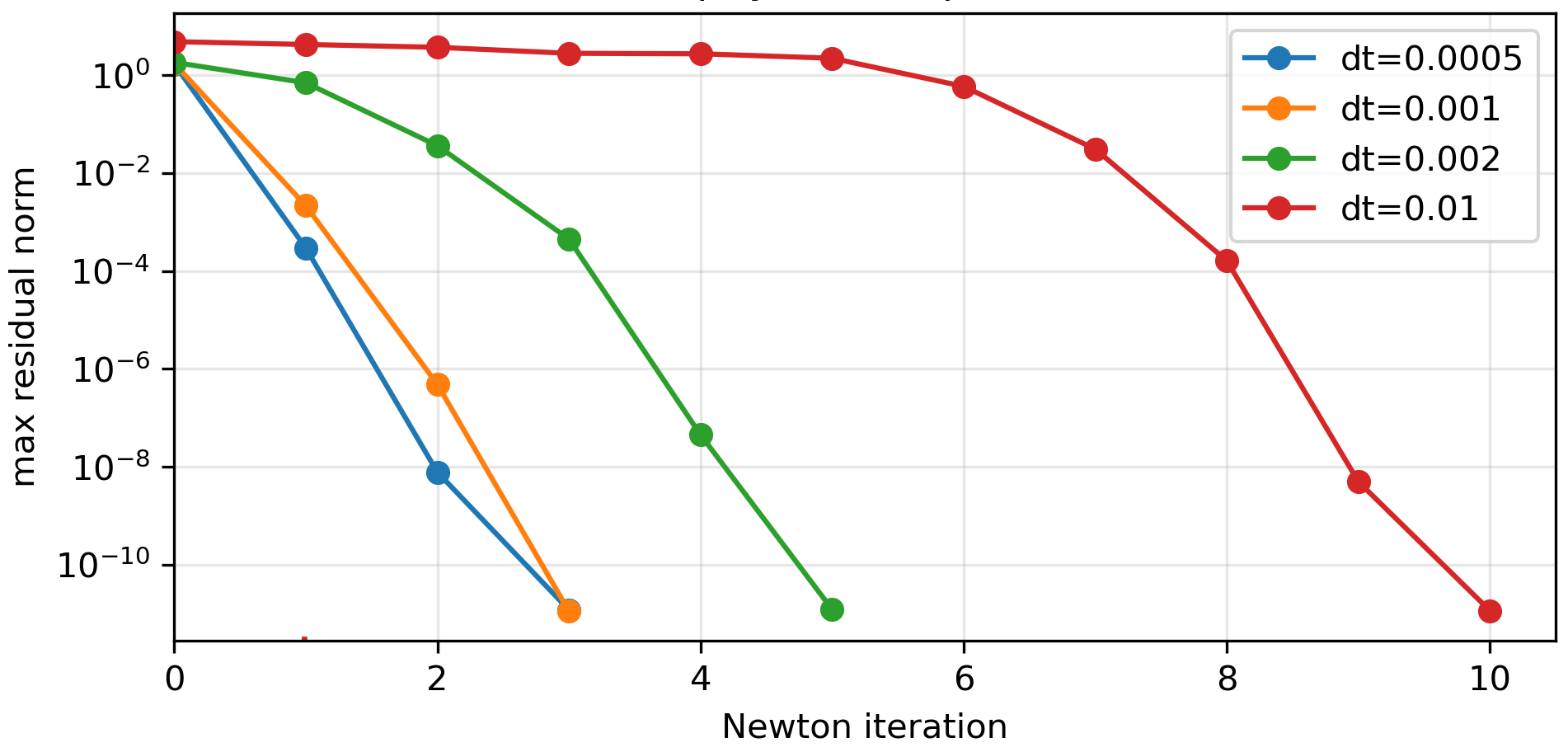}
  \caption{The maximum residual norms over all transport channels as a function of Newton iteration at the first and final time steps. Results are shown for two stiffness coefficients $a_{Te/Ti/ne}=(0.2, 0.2, 0.4)$ (left) and $a_{Te/Ti/ne}=(2, 2, 4)$ (right) and four physical time steps, $\Delta t=5\times10^{-4}$, $10^{-3}$, $2\times10^{-3}$, and $10^{-2}$. In all cases, the nonlinear iterations converge rapidly, and the smaller time steps exhibit faster reduction of residuals. The comparison between the first and fortieth physical steps indicates that the coupled solve remains robust after the profiles have evolved away from their initial conditions. The large stiffness makes convergence more difficult.}
  \label{fig:dd_converge}
\end{figure}

The results obtained in this test for coupled transport equations, as shown in Figs.~\ref{fig:dd_profiles} and \ref{fig:dd_converge}, demonstrate stable nonlinear convergence and physically correct redistribution for all three transport channels in the simulation that combine sources, diffusion, and cross-channel coupling. In this test case, both the electron and ion temperature equations are coupled, while the plasma density channel evolves independently. The input file for this test case uses $N_x=81$ radial points, $\Delta t=5\times10^{-4}$, $t_{\text{final}}=2\times10^{-2}$, and up to 100 Newton iterations per time step, where both residual and relative change tolerances are set to $10^{-8}$ and nonlinear stabilization is enabled. The electron and ion temperature channels are initialized using Gaussian profiles and use an off-axis Gaussian source term and stiff diffusion, while the plasma density channel is initialized using a Gaussian profile and uses a parabolic source term localized at the plasma edge and stiff diffusion. Zero gradient boundary conditions are used at the magnetic axis, while constant boundary conditions are set for the outer boundary with $T_{\text{e}}=0.6$, $T_{\text{i}}=0.5$, and $n=0.9$. The temperature coupling matrix only includes off-diagonal elements $c_{\text{ei}}=c_{\text{ie}}=8$. As a result, only electron–ion temperature relaxation is allowed without any coupling to plasma density. The profiles evolve smoothly into broader and less peaked shapes, and the strongest coupling effect is seen in the temperature channels. The convergence histories show monotonically decreasing maximum residual and relative change, as required for nonlinear convergence in PT\_SOLVER, and demonstrate the expected behavior that decreasing physical time steps require fewer and less severe Newton iterations for convergence.

As a further illustration, the sensitivity of nonlinear convergence to the physical time step and the stiffness in the gradient-dependent transport model is demonstrated in Fig. \ref{fig:dd_converge}. The residual is always reduced monotonically as the solver converges using the Newton iterations, and this is true in all cases considered here, at both the initial and final time steps. The effect of the stiffness is clearly demonstrated, as the reduced stiffness case shows faster convergence with a smaller number of iterations and a faster rate of residual reduction, whereas the higher stiffness case shows a slower rate of residual reduction and a greater number of iterations. The latter is consistent with the form of the model diffusivity in Eq.~\ref{eq:stiffmodel}, as the sensitivity of the effective transport to local gradients is increased as the stiffness is increased, thus making the problem more nonlinear in the implicit solver. The results confirm the PT\_SOLVER convergence in the full range of the parameter sweep, and they also confirm that higher stiffness has a significant impact on the nonlinear solver.

Although the theory-based model may be expected to be less prone to the effects of anomalous transport coefficients than the explicit stiffness model, the numerical problem related to the stiffness in the transport coupled equations still persists in the sense that the transport coefficients may be significantly influenced by the evolving profiles, and this is a problem that is common in a wide range of applications involving the simulation of the transport of reacting systems, as the transport coefficients may be significantly influenced by the evolving profiles in the reactions and the diffusion processes. The numerical problem is therefore not so much related to the explicitness of the stiffness, but rather whether the transport model itself is nonlinear in the solver.

For realistic predictive calculations, the nonlinear transport problem may be more complex than that treated in this idealized stiffness scan. As mentioned in this paper, theory-based transport models are generally functions of multiple and competing gradients rather than a single control parameter. For example, in the thermal transport driven by the ion temperature gradient modes, the destabilizing effect of the ion temperature gradient and the stabilizing effect of the density gradient are coupled in such a way that the rate of evolution to marginal stability is determined by a complex interplay between these two gradients rather than by the monotonically changing value of a single gradient. As a result, the stiffness of the transport response can vary significantly in both space and time and may change rapidly in response to changing profiles. Another complication in realistic TRANSP calculations is that not all profiles are self-consistently evolved in response to plasma evolution. Some profiles may be provided as inputs based on experimentally determined fits to analogous plasmas or external models. In stiff transport calculations, this noise is undesirable because it directly affects local gradients used in determining transport coefficients and may lead to undesirable oscillations in local diffusivities, which would degrade nonlinear convergence.

Beyond the parallel scaling strategy described above, PT\_SOLVER also addresses the practical cost of predictive simulations through the modular design of its transport-model interface. Because the solver advances the transport equations separately from the internal details of a particular transport model, the same framework can be coupled not only to standard reduced theory-based models but also to machine-learning surrogates of those models. This feature becomes especially crucial when the predictive simulations demand small physical time steps or extended pulse lengths, since repeated evaluations of the computationally costly transport models could become too expensive. In this sense, PT\_SOLVER provides a pathway for incorporating transport physics of different fidelity into integrated modeling workflows without sacrificing numerical robustness or practical turnaround time.

\subsection{Verification with the TGYRO code}

\begin{figure}[!b]
 \centering
  \includegraphics[width=0.95\textwidth]{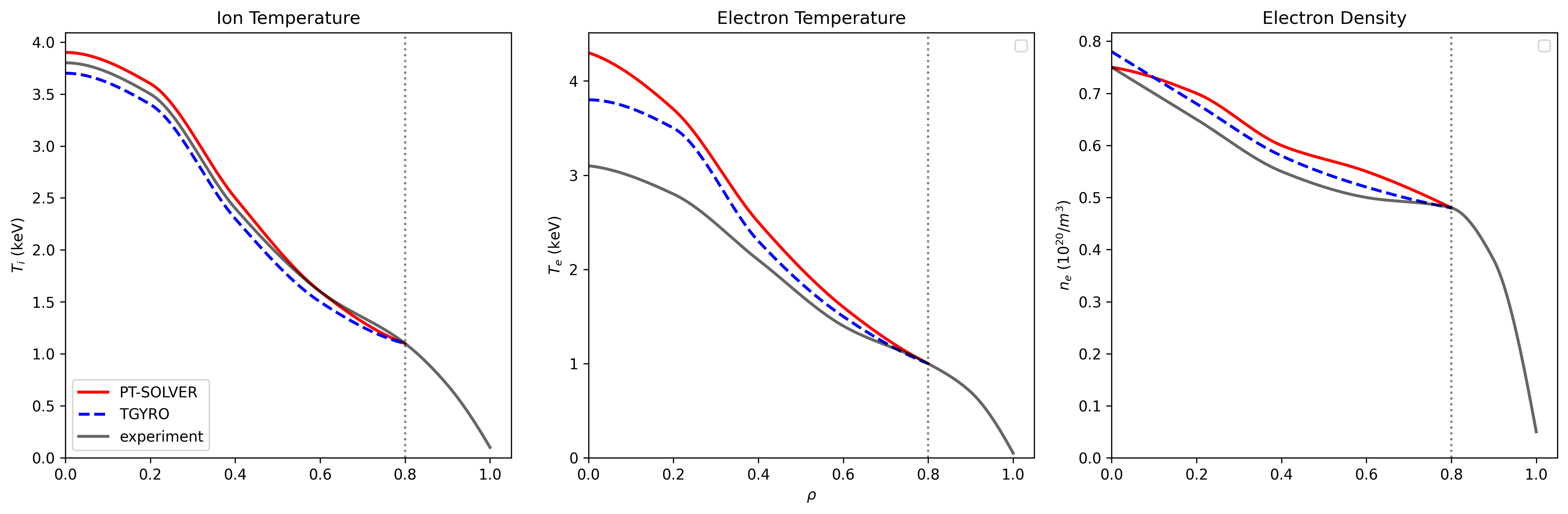}
  \caption{Comparison of standalone PT\_SOLVER and TGYRO results for the DIII-D H-mode discharge 125236 at 3.5~s using TGLF for anomalous transport and NEO for neoclassical transport in both workflows. Red solid curves denote PT\_SOLVER, blue dashed curves denote TGYRO, and gray curves denote the experimental profiles. The dotted vertical line at $\rho=0.8$ marks the outer boundary of the predictive region. The agreement is generally good for $T_i$ and $n_e$, while the largest difference is observed for $T_e$ in the inner core.}
  \label{fig:pt_d3d}
\end{figure}

The use of stiff turbulence models for cross-verification of transport solvers presents a challenge because the transport fluxes are determined based on local profile gradients as well as the specific details of the magnetic equilibrium, which include both the q-profile and magnetic shear. In addition to the gradients, these quantities can play a critical role in the transport level predicted with the models. Small differences in the numerical representation of equilibrium or from the interpolation of input data and sources can lead to significant differences in profile predictions and, in some cases, result in a different branch of instability driving the anomalous transport. Therefore, for the cross-verification of two predictive solvers, both solvers must be aligned as much as possible and use the same equilibrium, boundary conditions, and sources so that the two transport model inputs are consistent. Therefore, in order to limit any differences unrelated to the transport solver being verified (e.g., to limit differences between the two transport solvers being compared), both PT\_SOLVER and TGYRO use the equilibria reconstructed {\ap{with TEQ equilibrium solver in TRANSP}} rather than evolving them in a self-consistent manner.

{\ap{PT\_SOLVER advances the transport equations using face-centered effective diffusivities and convective velocities. This approach is compatible with the need for the implicit finite volume operator, even when the transport model, in this case, TGLF, provides total turbulent fluxes that cannot be uniquely decomposed into diffusive and convective contributions. As a result, after each call to the transport model, the total heat, particle, and momentum fluxes are transformed into their equivalent diffusive-convective form
\begin{equation}
q_s^{\rm turb}
=-n_s \chi_s^{\rm eff}\nabla T_s+n_s T_s V_{T_s}^{\rm eff},
\qquad
\Gamma_s^{\rm turb}
=-D_s^{\rm eff}\nabla n_s+n_s V_{n_s}^{\rm eff},
\end{equation}
where \(q_s^{\rm turb}\) is a heat flux density and
\(\Gamma_s^{\rm turb}\) is a particle flux density. For toroidal angular
momentum,
\begin{equation}
\Pi_\phi^{\rm turb}=-I_\phi \chi_\phi^{\rm eff}\nabla\Omega_\phi
+I_\phi\Omega_\phi V_\phi^{\rm eff}.
\end{equation}
In practice, the effective diffusivity is obtained from the part of the
model flux proportional to the local profile gradient, for example
\begin{equation}
\chi_s^{\rm eff}=-\frac{q_s^{\rm turb}}{n_s\,\widetilde{\nabla T_s}},
\qquad
D_s^{\rm eff}=-\frac{\Gamma_s^{\rm turb}}
{\widetilde{\nabla n_s}},
\qquad
\chi_\phi^{\rm eff}=-\frac{\Pi_\phi^{\rm turb}}
{I_\phi\,\widetilde{\nabla\Omega_\phi}},
\end{equation}
where the tildes denote the regularized gradients used in the numerical
implementation to avoid division by very small local gradients. The
remaining part of the total flux is represented as an effective convective
velocity,
\begin{equation}
V_{T_s}^{\rm eff}=\frac{q_s^{\rm turb}+n_s\chi_s^{\rm eff}\nabla T_s}
{n_sT_s},
\qquad
V_{n_s}^{\rm eff}=\frac{\Gamma_s^{\rm turb}+D_s^{\rm eff}\nabla n_s}
{n_s},
\end{equation}
with an analogous expression for \(V_\phi^{\rm eff}\). This decomposition should not be interpreted as denoting an actual physical dissociation between the two contributions to the total turbulent flux. Rather, it serves as a way of numerically reconstructing the total turbulent flux in such a way that the diffusive-convective form, which is used by the PT\_SOLVER transport operator, may be obtained from the total turbulent flux calculated by TGLF model. The effective transport coefficients are then re-evaluated at each iteration step based on the profiles available in the solution. Because the reconstruction depends on local gradients, small differences
in profile interpolation or gradient evaluation can change the effective
\(\chi\)--\(V\) split without necessarily implying a corresponding change
in the total model flux.}}

{\ap{Many discharges have been studied previously with the GYRO, TGYRO/TGLF, and TRANSP~\cite{Kinsey2011,Wang2010}. One example is the  DIII-D discharge 125236, an H-mode with 1.1 MA plasma current, 2.1 T toroidal field, and heated with 5.5 MW neutral beams~\cite{Solomon2007,Budny2012}.}} PT\_SOLVER and TGYRO predict similar transport behavior for the majority of their predictive region from $\rho=0$ to $\rho=0.8$ where the boundary conditions were set. Both codes use the TGLF model~\cite{Staebler2005TGLF,Kinsey2008TGLF,Staebler2007TGLF} for the anomalous transport and the NEO model~\cite{Belli2008KineticCO} for the neoclassical transport. Experimental profiles are included for reference. The vertical dotted line at $\rho=0.8$ represents the edge of the predictive region. {\ap{The differences in TGYRO and PT\_SOLVER settings are summarized in Table~\ref{tab:ptsolver_tgyro}.}}

\begin{table}[t]
\centering
{\ap{\caption{
Summary of the matched PT\_SOLVER and TGYRO setup used for the DIII-D 125236 verification case. The purpose of this comparison is solver and workflow verification, not validation of the underlying transport model against experiment. Both workflows used the same experimental equilibrium, boundary conditions, sources, and transport-model settings as closely as possible. }
\label{tab:ptsolver_tgyro}
\begin{tabular}{p{0.20\textwidth}||p{0.36\textwidth}|p{0.36\textwidth}}
\hline
\textbf{Quantity} & \textbf{PT-SOLVER / TRANSP} & \textbf{TGYRO} \\
\hline
\hline
Numerical algorithm &
Time-dependent implicit profile advance to a quasi-stationary solution
using the TRANSP predictive workflow &
Steady-state flux-matching calculation using the same transport and
source inputs as closely as possible \\
\hline
Radial coordinate &
TRANSP normalized toroidal-flux coordinate, \(\rho = \sqrt{\Phi/\Phi_{\rm sep}}\) &
Same normalized radial coordinate after mapping from TRANSP \\
\hline
Predictive domain & 
\(0 \leq \rho \leq 0.8\);  &
Same predictive domain  \\
\hline
Radial grid &
Native uniform TRANSP radial grid with \(N_{\rho}^{\rm PT}=100\) &
TGYRO grid samples the same predictive region \\
\hline
Evolved channels &
\(T_e\), \(T_i\), \(n_e\), and $v_{\rm phi}$ &
Same evolved channels. \\
\hline
Equilibrium treatment &
Experimental reconstructed equilibrium with the TEQ solver in TRANSP held fixed during the comparison. &
Same fixed equilibrium. \\
\hline
Electron response & Non-adiabatic electrons. & Same electron treatment. \\ 
\hline
Electromagnetic effects & 
\multicolumn{2}{p{0.74\textwidth}}{Finite-\(\beta\) electromagnetic effects are included through the perpendicular magnetic perturbation contribution. The parallel magnetic perturbation contribution is not included. } \\ 
\hline
\(E\times B\) and pressure-gradient stabilization & 
\multicolumn{2}{p{0.74\textwidth}}{ \(E\times B\) shear stabilization is included using the rotation and equilibrium information from the matched TRANSP setup. Pressure-gradient stabilization and finite-\(\beta\) electromagnetic effects are retained in the turbulent transport calculation. No additional user rescaling or quench correction is applied to these stabilization terms. } \\
\hline
Heating and torque sources & 
\multicolumn{2}{p{0.74\textwidth}}{ Heating and torque sources are taken from the matched TRANSP setup. They are primarily due to deuterium neutral beam injection computed with NUBEAM. Ohmic heating and radiative losses are included, while no active RF heating or current-drive model is used. } \\ 
\hline
Particle sources & 
\multicolumn{2}{p{0.74\textwidth}}{ Particle sources include neutral beam deposition, gas fueling, recycling, and thermal-neutral sources from TRANSP. Pellet fueling is not enabled. The carbon impurity model contributes to dilution, \(Z_{\rm eff}\), and radiation. } \\
\hline
Impurities &
\multicolumn{2}{p{0.74\textwidth}}{A carbon impurity species is included, with \(Z=6\) and \(A=12\). The impurity content and effective-charge information are taken from the TRANSP input data.} \\
\hline
Boundary condition & \multicolumn{2}{p{0.72\textwidth}}{Boundary values from experiment at \(\rho=0.8\).} \\
\hline
\end{tabular}}}
\end{table}
\FloatBarrier

{\ap{To quantify the comparison shown in Fig.~\ref{fig:pt_d3d}, the PT\_SOLVER and TGYRO profiles are interpolated onto identical radial points and compared in the common predictive range of \(0 \leq \rho \leq 0.8\). Consistent with the setup summarized in Table~\ref{tab:ptsolver_tgyro},  this comparison is not aimed at validating the physics-based models, such as the TGLF model and the NEO models, or verifying predictions using experimental data. Rather, the goal is to verify that PT\_SOLVER provides results comparable to the solution from another transport code with the same transport models, identical equilibrium state, and matching sources. In particular, the saturation-rule setting is included for reproducibility, but this
comparison is not intended to determine the preferred TGLF saturation
rule for this discharge; the experimental profiles are shown only as
context for the realistic plasma state used in the solver comparison.}}

{\ap{The normalized RMS error and maximum pointwise relative error are given as follows:
\begin{equation}
    \delta f_{\rm RMS} =
\left[\frac{\sum_j w_j
\left(f^{\rm PT}_j-f^{\rm TGYRO}_j\right)^2}
{\sum_j w_j\left(f^{\rm TGYRO}_j\right)^2}
\right]^{1/2},
\qquad
\delta f_{\max}
= \max_j \left| \frac{f^{\rm PT}_j-f^{\rm TGYRO}_j} {f^{\rm TGYRO}_j} \right|,
\end{equation}
where \(f\in\{T_i,T_e,n_e\}\), and \(w_j=1\) was used for the present radial-grid comparison. The resulting differences are
\[ \begin{array}{c|cc} \hline \text{Profile} & \delta f_{\rm RMS} & \delta f_{\max} \\ \hline T_i & 6.4\% & 5.7\% \\ T_e & 8.4\% & 13.2\% \\ n_e & 3.5\% & 3.9\% \\ \hline \end{array} \] 
The comparison should not be interpreted as requiring exact convergence to the same steady-state fixed point. TGYRO provides a solution to the flux-matching steady state problem, while PT\_SOLVER resolves time-dependent transport equation problems using implicit time stepping and finite time convergence. Hence, what needs to be verified here is whether the time-dependent PT\_SOLVER process provides solutions for profiles that are consistent with the steady-state transport process using the same realistic transport model, equilibrium, boundary condition, and source profiles. As far as this requirement is concerned, the match that has been achieved so far is sufficient, and it shows that PT\_SOLVER handles realistic stiff transport closure in a similar way as some other solver used in practice. Some deviations can be expected, however, because even small errors in the mapping, interpolation, or especially gradient construction could be highly amplified by stiff transport, where flux strongly depends on local gradients.}}

The largest differences between solvers are observed for the electron temperature in the core, while the ion temperature and the electron density are more consistent. It appears that both solvers predict the electron temperatures above the experimental values. Since this benchmark uses a common transport-model setup but is not fully self-consistent, the over-prediction of $T_e$ relative to experiment should be attributed to the shared model configuration rather than to differences between the two solvers. The remaining difference between PT\_SOLVER and TGYRO is plausibly related to differences in solver formulation, including the time-dependent evolution in PT\_SOLVER versus the steady-state flux-matching approach in TGYRO, together with differences in source interpolation and other workflow details. However, despite the differences in the electron temperature predictions from PT\_SOLVER and TGYRO, they are still in agreement with each other for most part of the profile.

For the purposes of the present paper, these results should be regarded as solver and workflow verification rather than comprehensive model validation.  Specifically, this research does not have an objective to establish predictive accuracy or fidelity using experimental data, so there is no way to evaluate whether the TGLF and NEO predictions are accurate based on such tests. At the same time, recent database-oriented studies, including DIII-D, MAST-U, and NSTX comparisons~\cite{Abbate2024,Lestz2025_AssessingNSTX,Lestz2025_MMMSensitivityNSTX,Neiser2024TTF}, have addressed validation of related reduced-model and predictive workflows and provide important complementary context, but those studies are outside the scope of the present paper. The analysis in this paper shows that the profile predictions produced by PT\_SOLVER are consistent with the TGYRO results, provided that the equilibrium, source terms, and boundary conditions are controlled as closely as possible. This verification test supports the conclusion that PT\_SOLVER is numerically consistent with the reference workflow for the matched TGLF/NEO benchmark configuration.

\subsection{Multilevel parallelization strategy and scaling}
\begin{wrapfigure}{r}{0.42\textwidth} 
\vspace{-20pt}
  \begin{center}  \includegraphics[width=0.42\textwidth]{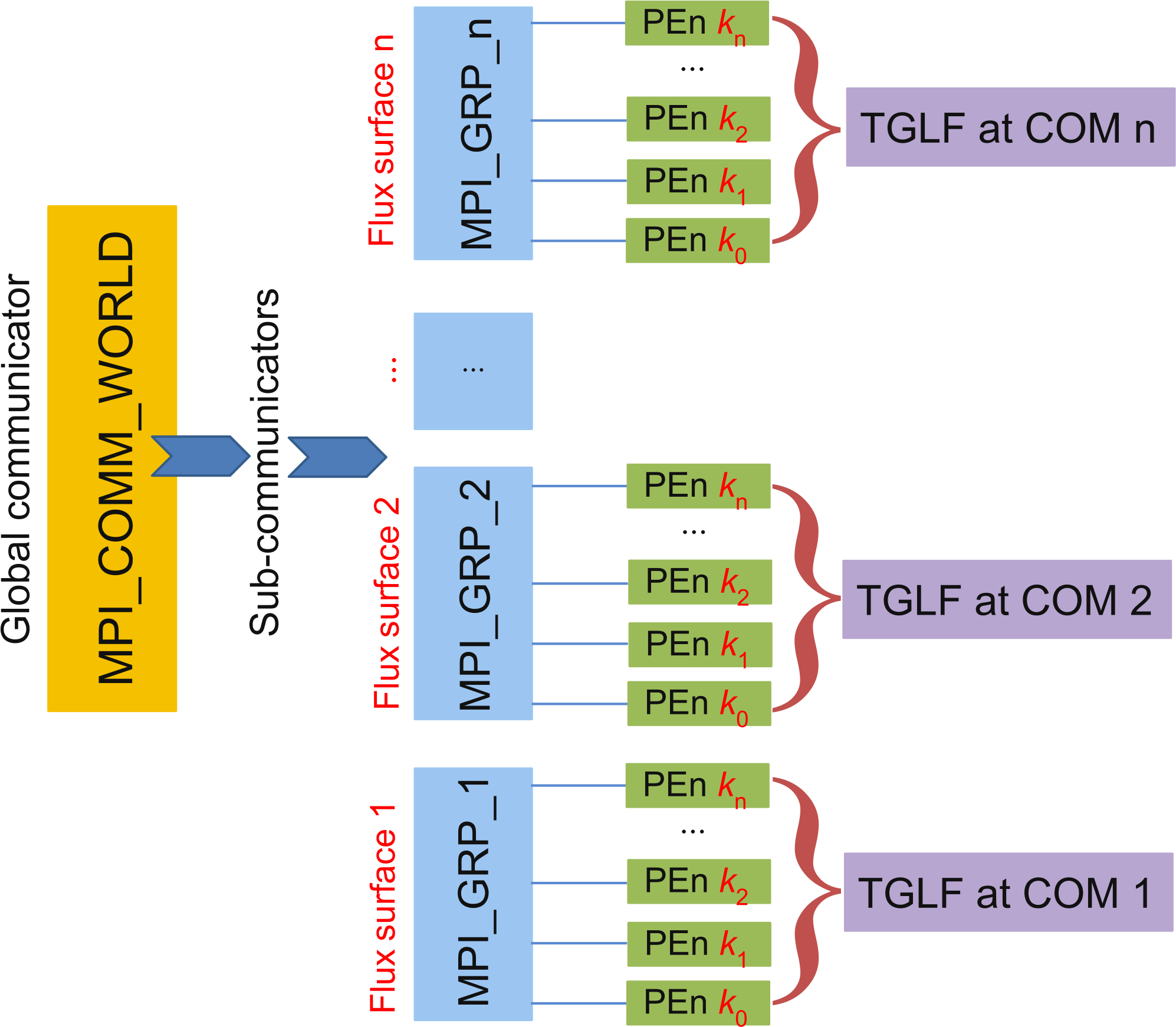}\vspace{-10pt}
  \caption{Schematic illustration of the multilevel parallelization strategy used in PT\_SOLVER for coupled transport calculations with TGLF.\vspace{-15pt}}
  \label{fig:pt_parallel}
  \end{center}
\end{wrapfigure}

for
PT\_SOLVER is designed as a parallel modular solver with multiple levels of parallelization. The nonlinear profile advance and the expensive transport-model evaluations can be decomposed across multiple levels of parallel communication. The MPI\_COMM\_WORLD communication pattern shown in the diagram in Fig.~\ref{fig:pt_parallel} has been divided into multiple sub-communicators, one per flux surface. The implementation is shown on an example of the TGLF model that maintains its own parallelization across different wavenumbers. The global communicator, MPI\_COMM\_WORLD, is partitioned into sub-communicators, MPI\_GRP\_$i$, associated with individual radial flux surfaces. Within each sub-communicator, processors are assigned to the parallel evaluation of TGLF over wavenumber space, denoted here by $k_0,\ldots,k_n$. These calculations are associated to individual PEn, or $n$-th MPI process ranks. This two-level decomposition enables simultaneous parallelization over flux surfaces and over the internal TGLF calculation at each surface. The transport coefficients computed using these transport model calculations are collected and passed back to the Newton iteration for the transport equation advancement. The resulting architecture enables scalable processor use beyond just being limited to the number of radial zones, thereby increasing the effective use of both radial and $k_y$ spatial concurrency.

\begin{figure}[!b]
 \centering
  \includegraphics[width=0.95\textwidth]{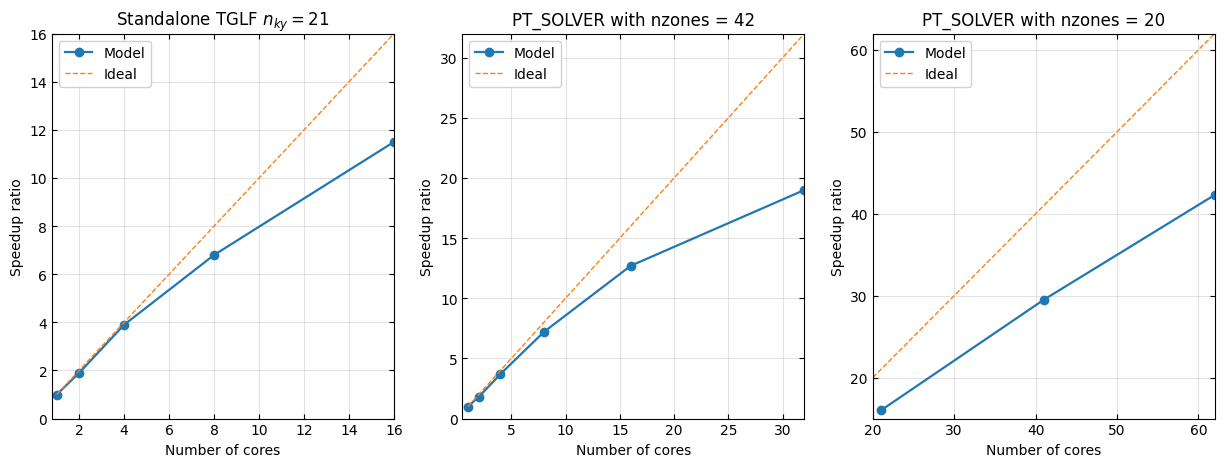}
  \caption{Parallel scaling of the PT\_SOLVER with the TGLF model for three representative configurations. Left: scaling of standalone TGLF with $n_{ky}=21$. Middle: scaling of PT\_SOLVER TGLF for $n_{\mathrm{zones}}=42>N_{cores}$. Right: scaling of PT\_SOLVER with parallel TGLF for $n_{\mathrm{zones}}=20<N_{cores}$. In each panel, the solid curve shows the measured model speedup, and the dashed line indicates ideal linear scaling. Standalone TGLF exhibits good scaling up to the number of processors comparable to the number of $k_y$ modes, while PT\_SOLVER benefits from an additional level of parallelism over transport zones. When both levels of parallelism are used simultaneously, the combined workflow maintains substantial speedup, although with the expected departure from ideal scaling due to communication and load-balance overheads.}
  \label{fig:pt_scaling1}
\end{figure}

The scaling results that correspond to this implementation are shown in Fig.~\ref{fig:pt_scaling1}. The first scaling data set indicates that TGLF has a reasonable scaling up to the limits of $n_{ky}$ decomposition. The second plot in Fig.~\ref{fig:pt_scaling1} corresponds to the conditions where the number of zones is greater than the number of CPUs so that PT\_SOLVER and TGLF operate in parallel; however, PT\_SOLVER creates only one sub-communication group, and internal parallelization over wavenumbers inside TGLF is used. The third scaling plot in Fig.~\ref{fig:pt_scaling1} indicates a condition where the number of zones is less than the number of CPUs, so that both PT\_SOLVER and TGLF are performing in parallel mode. The three scaling examples demonstrate the automatic transition of the two workflow modes of operation for PT\_SOLVER and TGLF: PT\_SOLVER operating in parallel mode and TGLF operating in serial mode, when the radial workload is primarily located within one or more transport zones, and both PT\_SOLVER and TGLF operating in parallel mode, when additional CPUs are available beyond the total number of zones in the transport; thus allowing for efficient use of available CPU resources.

The scope of this PT\_SOLVER verification should be limited to the presented benchmark hierarchy. The objective of this verification is to confirm the numerical convergence of the predictive solver, and the correctness of the interface implementation and the associated workflow, not to provide a comprehensive validation of all transport models available in TRANSP.  In particular, the simplified gradient-dependent diffusivity tests are used solely to isolate the stiffness and convergence characteristics of the nonlinear solver, while the TGYRO comparisons are meant to confirm that both the solver and the workflow are consistent with TGLF and NEO inputs. However, we are not evaluating the predictive accuracy compared to experimental data. The multilevel scaling findings are an indication of the PT\_SOLVER/TGLF interface and must not be used as a general performance metric for all model combinations. The verification here focuses specifically on the channels and workflow that are exercised in the benchmark set to show how robust the current numerical schema is under these conditions. Overall validation, along with additional evaluations of frameworks, will be conducted in the future.

\section{Implementation of T3D/GX in TRANSP for Gyrokinetic Turbulence Simulations\label{sec:gx}}
T3D is a transport solver developed in Python and designed to perform self-consistent macro-scale plasma profile evolution in a modular fashion, while the GX code represents a nonlinear gyrokinetic turbulence code designed to provide high-fidelity turbulent flux calculations on GPU architectures. The goal of this section is not to assert the physics validity of the coupled TRANSP/T3D/GX process. Instead, it is to describe the coupling process and to highlight the state information exchanged between the two codes and the verification process to ensure the correctness of the interface implementation. This is consistent with the verification of solvers presented in this work, where the focus is on the numerical correctness and integrity of the coupled process rather than the transport process itself, and comparison to experiment.

The justification behind connecting TRANSP to T3D/GX is not to replace the reduced theory-based models available through PT\_SOLVER. Rather, it is a step toward developing a flexible approach where selected cases using reduced and high-fidelity models can be compared within the same modeling environment, using a common framework with the same interfaces and data structures. This capability is valuable for interface verification, assessment of reduced-model assumptions, and future hybrid workflows in which expensive turbulence calculations may be combined with surrogate or reduced descriptions for practical predictive applications.

The background and implementation of the GX code follow the Laguerre–Hermite representation developed by Mandell \textit{et al.}, who showed that this representation provides a smooth interpolation between low-resolution gyrofluid-like and high-resolution gyrokinetic limits \cite{Mandell2018}. The recent publication of the current version of the GX code describes it as a gyrokinetic turbulence code designed to run on GPU architectures and describes the numerical design choices that provide rapid nonlinear flux calculations on a single or few GPUs \cite{Mandell2024}. The T3D code follows the multiscale paradigm of direct multiscale coupling between transport evolution and local turbulence calculations established in the TRINITY framework \cite{Barnes2010}. The recent flux-driven simulations using T3D in the context of recent predictive simulations, including STEP-oriented applications, also highlight the relevance of the T3D code in transport simulations rather than as a comparison code \cite{Giacomin2025}. 

\subsection{Integration in TRANSP}

To minimize disruption to the existing predictive framework, the T3D/GX integration in TRANSP was organized around a narrow and explicit interface. TRANSP remains responsible for the integrated workflow: equilibrium handling, source calculation, boundary conditions, and top-level predictive control. The coupling layer exports geometry and state information, launches T3D with template-based inputs, and re-imports the evolved profiles. In practical terms, the operational interface follows the documented TRANSP coupled-run workflow in which \texttt{LPREDICTIVE\_MODE=5} activates the T3D pathway and the run directory contains a TRANSP namelist together with inputs for the T3D (\texttt{T3D\_template.in}) and GX (\texttt{GX\_template.in}) codes \cite{transpT3DGXGuide}. The official guide also makes explicit that the T3D import mode for the coupled branch is \texttt{transp\_realtime} \cite{transpT3DGXGuide}.

\begin{figure}
\includegraphics[width=0.96\linewidth]{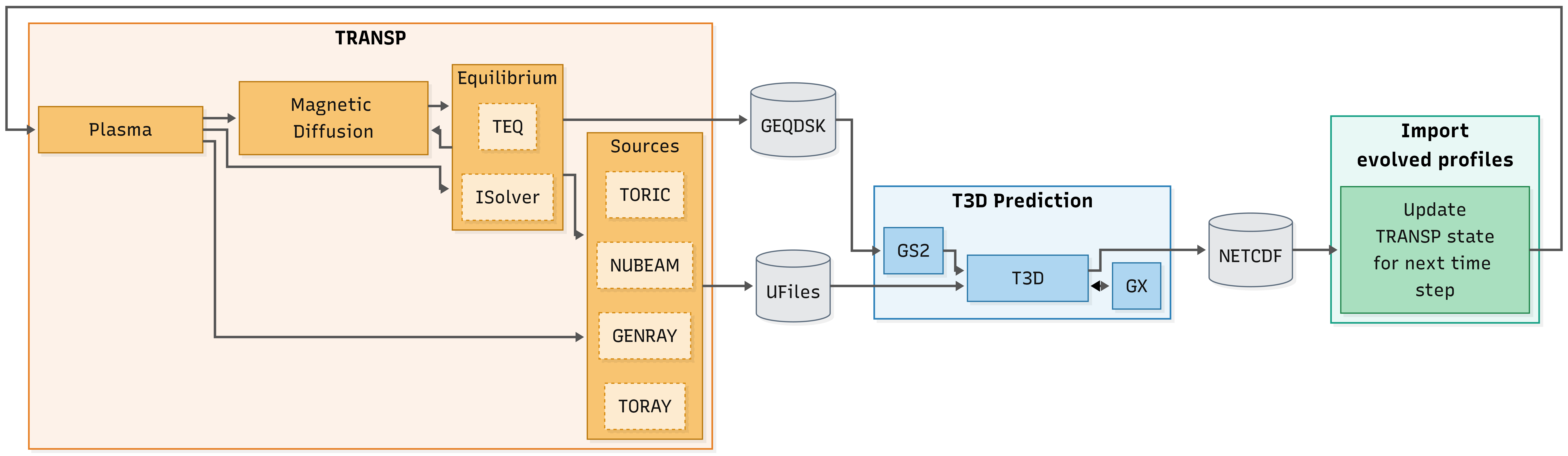}
\vspace{0.3em}
\caption{Interface-focused schematic of the TRANSP--T3D/GX coupling shows the modular organization and the direction of information exchange. \label{fig:integration}}
\end{figure}

A schematic representation of the TRANSP-T3D interface is shown in Fig.~\ref{fig:integration}. The coupling uses exported geometry and state objects -- specifically GEQDSK-like geometry, profile and source data, and template-driven run control -- to keep the external solver pathway synchronized with the predictive TRANSP state. This choice is also consistent with the broader TRANSP philosophy of interfacing external components through explicit, traceable data objects rather than by duplicating physics logic in several locations \cite{APankin2024}.

In the coupled density-prediction branch, charge neutrality and effective charge are enforced through
\begin{align}
n_e &= \sum_i n_i + \sum_{j,q} Z_q n_j^q, \\
n_e Z_{\mathrm{eff}} &= \sum_i n_i + \sum_{j,q} Z_q^2 n_j^q,
\end{align}
where the index $i$ denotes singly charged main-ion species and $n_j^q$ denotes the density of impurity species $j$ in charge state $q$. These relations are applied during density prediction so that, for example, if electron density is evolved, the total thermal-ion density and impurity content can be reconstructed consistently from the known $Z_{\mathrm{eff}}$ while the fast-ion contribution is held fixed over the transport step. This makes explicit where consistency is enforced in the interface rather than leaving a reviewer to infer that the coupled branch might duplicate or silently modify these constraints. It also parallels the predictive-TRANSP statement that charge neutrality and $Z_{\mathrm{eff}}$ constrain density evolution in the native solver path \cite{APankin2024}.

A second point that deserves explicit clarification is the treatment of collisional electron--ion energy exchange. In predictive TRANSP/PT\_SOLVER, collisional exchange appears as the term $Q_{ei}$ in the electron and ion power-balance equations and enters the two channels with opposite signs, acting as a direct source for one species and an equal sink for the other \cite{APankin2024}. In T3D, the same physical process is represented in the species pressure equations through a collisional relaxation operator of the form
\begin{equation}
\frac{\partial p_s}{\partial t}
+\frac{1}{V'}\frac{\partial}{\partial \Psi}\!\left(V' \overline{Q_s}\right)
=
\frac{3}{2} n_s \sum_u \nu_{su}\left(T_u-T_s\right)
+ S_{p,s},
\end{equation}
where the sum is over collisional partners $u$ and the exchange coefficient is represented through $\nu_{su}$ \cite{Barnes2010,Giacomin2025}. Thus, the coupled benchmarks do not compare two literally identical algebraic implementations. Instead, they verify that the interface maps the same underlying inter-species relaxation physics consistently between the native TRANSP formulation and the T3D formulation. This distinction is important because it explains why the $Q_{ei}=0$ cases isolate pure interface bookkeeping, whereas the $Q_{ei}\neq 0$ cases additionally test consistency of the inter-channel coupling logic.

\subsection{Verification of the TRANSP/T3D interface}

\begin{figure}[htbp]
    \centering
    \includegraphics[width=.98\textwidth]{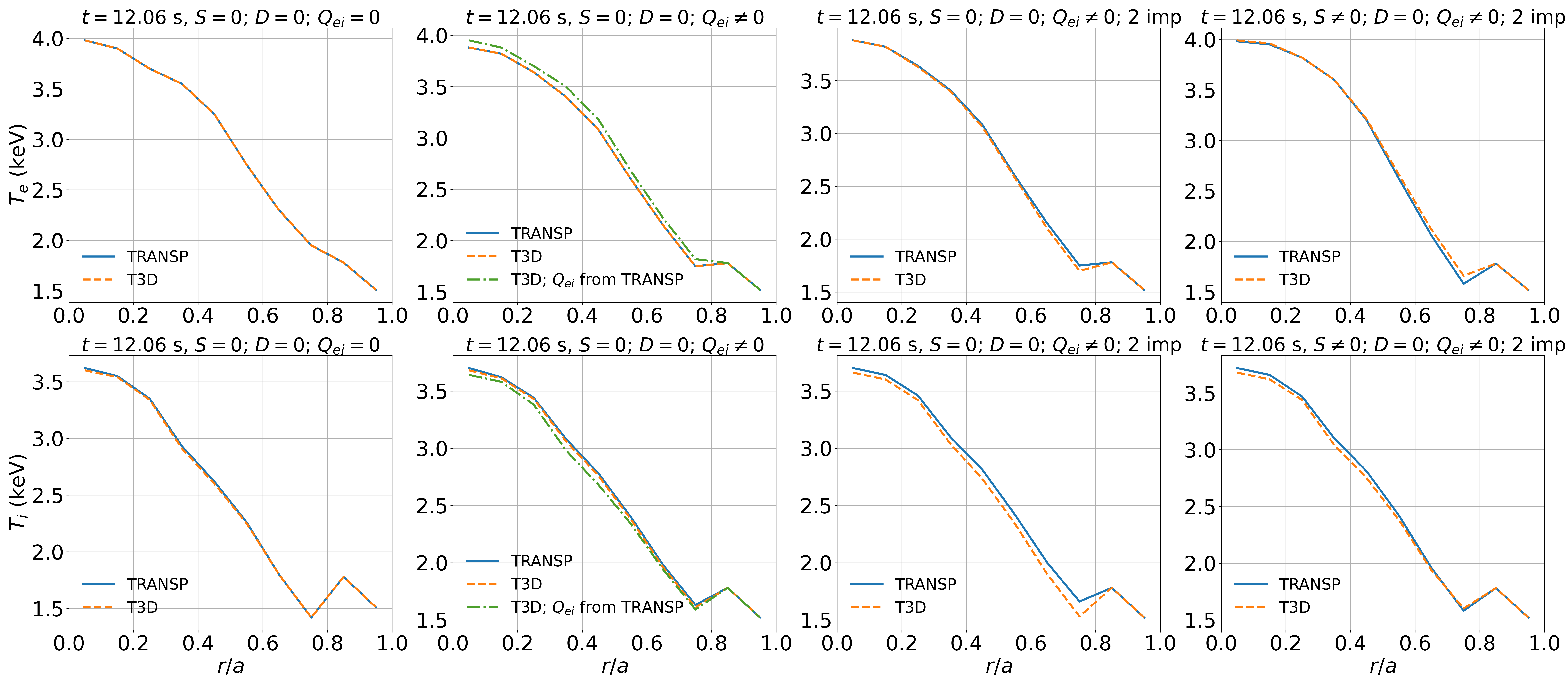}\vspace{-2mm}
    \caption{{\ap{Comparison between predicted profiles using PT\_SOLVER and T3D in different controlled interface verification runs. The case being considered is the JET discharge 42982. No ion temperature measurements in the plasma edge region were available, so the electron temperature determines the ion temperature boundary condition in this case. This introduced the artificial discontinuity in $T_i$, used here as a practical way to indicate collisional coupling. The tests shown in the table become increasingly complex going from left to right. Here, $S$ stands for source, $D$ represents the diffusion/convective contribution from the source term, $Q_{ei}$ is collisional electron-ion energy exchange, while ``2 imp.'' means plasma with carbon and beryllium impurities. This test set aims at verifying profile mapping, source terms, impurity accounting, and electron-ion collisional coupling at the interface between TRANSP and T3D; they do not produce self-consistent GX-driven fluxes.}}}
    \label{fig:pt_t3d_comparison}
\end{figure}

The verification benchmark retained here is intended to answer a limited but important question: when T3D is connected through the TRANSP interface and supplied with matched inputs, does the coupled interface reproduce the same profile evolution expected from the predictive TRANSP formulation for the benchmark considered? The answer from the present test set is yes, within the accuracy visible in the benchmark panels shown in Fig.~\ref{fig:pt_t3d_comparison}. The benchmark results increase physics complexity in a controlled way. The first row removes auxiliary heating, diffusion, and collisional electron-ion exchange. The next rows introduce impurity content and $Q_{ei}$ coupling, and the final row adds a source term while retaining the impurity and collisional structure. This progression is useful because it tests not only the transport advance itself but also the interface bookkeeping associated with source terms, impurity-aware density constraints, and collisional exchange.

The first two columns show that the source-free cases preserve close agreement as impurity complexity and collisional exchange are introduced. The third column extends that check to the two-impurity case. The final column, which includes a nonzero source term, verifies that the coupled interface remains consistent once source handling is included. 

The benchmark shown in Fig.~\ref{fig:pt_t3d_comparison} is based on the JET discharge 42982, which was used previously in TRINITY verification works{\ap{~\cite{Barnes2010} and analyzed with TRANSP~\cite{Horton1999,Budny2000}. This was a historic plasma discharge performed during the 1997 DTE1 (Deuterium-Tritium Experimental 1) campaign. It produced the highest fusion power (4.7MW) in a baseline H-mode in that campaign~\cite{Zotta2022}.}} Ion-temperature edge data was not available for that discharge. Therefore, as is common practice in such situations, electron temperature is used as the basis for imposing the ion-temperature boundary condition. As a result, there appears an artificial discontinuity of the initial ion-temperature distribution at the predictive boundary. While this discontinuity is not physical, it is useful to verify our work as it provides us with an observable measure of the coupling between electron and ion channels through collisions. If $Q_{ei} = 0$, as seen in the first row of Fig.~\ref{fig:pt_t3d_comparison}, the discontinuity appears only in the ion channel and no effect is visible in the electron channel. If the collisions are turned on, however, as seen in the other figures, the corresponding effect appears in both ion and electron channels.

{\ap{The setup-up of particular benchmark shown in Fig. \ref{fig:pt_t3d_comparison} is chosen in order to reduce the errors due to radial-grid mapping between the coupled codes. For the TRANSP code, a total of 10 radial zones has been used, with the outer boundary being located at $\rho=0.85$, where $\rho$ stands for the normalized toroidal flux coordinate used in this benchmark. The T3D region has been restricted to $0.05 \leq \rho \leq 0.85$, while the GX flux-tube locations have been made to coincide with the TRANSP radial mesh. The goal of this setup has been to facilitate an interface verification test based on differences between radial meshes that will arise during interpolation.}}

{\ap{The examples shown in Fig. \ref{fig:pt_t3d_comparison} do not represent self-consistent GX-flux-driven transport simulations. Instead, they can be considered controlled interface tests with incrementally increasing physics content. The most simple case does not include any sources, prescribed transport contributions, or collisions of electron and ion populations. Additional cases include electron-ion collisions, source and prescribed transport terms. Finally, impurity content has been introduced into the latter case. }}

Since the largest differences are likely associated with differences in the implementation of collisional energy exchange in PT\_SOLVER and T3D, we briefly summarize those implementations in PT\_SOLVER and T3D here. In the T3D code, collisional energy exchange is treated using a fully multi-species formulation based on the fundamental expression 
\begin{equation}
 \frac{3}{2} n_s \sum_u \nu_{su} \left( T_u - T_s \right)
\end{equation}
In this approach, the energy transfer for each species $s$ is computed by explicitly summing over interactions with all other species $u$, with the coupling strength determined by the interspecies collision frequencies $\nu_{su}$. This formulation naturally captures electron–ion, ion–ion, and impurity interactions within a unified framework and retains the detailed pairwise collisional physics. As such, it represents a more fundamental, kinetic-based description of energy exchange processes in a multi-component plasma.

In contrast, the TRANSP code employs a reduced model for electron–ion heat exchange, in which the detailed multi-species interactions are approximated through a lumped parameter. Specifically, the ion contribution is represented by a charge–mass weighted ion density of the form $$\sum_s n_s \frac{Z_s^2}{A_s},$$ which enters the electron–ion coupling coefficient. This formulation is restricted to electron–ion interactions and avoids explicit summation over individual ion species by combining their effects into a single effective quantity. The weighting reflects the underlying collisional physics, with the $Z_s^2$ dependence enhancing the contribution of higher-charge ions and the $1/A_s$ factor accounting for the more efficient energy exchange associated with lighter ions. Consequently, TRANSP replaces the detailed, species-resolved collisional structure with an effective ion density that characterizes the aggregate coupling between electrons and the ion population.

Taken together, the benchmark Fig.~\ref{fig:pt_t3d_comparison} supports three specific claims.  First, the interface preserves the expected mapping of predicted channels and imported channels across the coupled pathway. Second, the constraint logic required for density prediction is being applied consistently when impurities are present. Third, the treatment of collisional exchange and source bookkeeping does not introduce spurious differences between the two solver paths over the cases considered. What these panels do not demonstrate is broad transport-model validation; that would require a separate study against experimental observables or against underlying turbulence calculations over a broader database. However, PT\_SOLVER in TRANSP and GX code in T3D, as well as GS2 code in TRINITY, have been verified before, and in this work, we verify only the numerical implementation of the interfaces. In other words, the present evidence should be read as interface verification and workflow verification, not as a general validation of T3D or GX physics performance.

The distinction matters for the scope of this paper. The present work is centered on predictive infrastructure in TRANSP and on the robustness of modular transport workflows. In that context, the TRANSP/T3D benchmark is valuable because it shows that the predictive framework can be extended in a controlled and traceable way to an external transport solver and, through it, to a high-fidelity turbulence model. This creates a practical path for future coupled studies with GX while keeping the present claims aligned with the evidence available.

\subsection{Performance Considerations}

One of the primary bottlenecks in these simulations arises from the computational intensity of resolving nonlinear gyrokinetic turbulence. These calculations necessitate high-resolution spatial and velocity grids, significantly increasing the numerical workload. Additionally, achieving accurate turbulence evolution requires fine temporal resolution, which further increases the computational burden. As a result, even short-duration simulations demand substantial computational resources, making predictive studies difficult to execute within practical time constraints.

Another limiting factor is the available GPU memory, which imposes constraints on the scale of turbulence calculations. Large-scale turbulence simulations require substantial memory allocations, and the need to store multiple field quantities, distribution functions, and intermediate results can lead to bottlenecks when working within the limitations of current GPU architectures. These memory constraints can restrict the spatial resolution and physics fidelity of the simulations, potentially limiting the accuracy of long-duration transport predictions.

Although linear simulations for stability analyses are significantly more computationally efficient, allowing them to be employed effectively for interpretive studies, predictive simulations remain computationally expensive. The high computational cost prevents their routine application in between-shot analysis, where rapid turnaround times are necessary. Addressing these performance limitations will require further optimizations in GPU implementation, potential algorithmic improvements, and the development of hybrid modeling strategies that balance fidelity and efficiency.

\subsection{Future Improvements}

Making predictive simulations using T3D/GX more feasible demands a set of strategic enhancements through the reduction of computational costs and improving efficiency. Part of this development entails optimization of the GX implementation on the GPU. As GX will be used for nonlinear gyrokinetic turbulent calculation, optimizing performance on GPU architecture is critical. Memory management optimization, kernel performance optimization, and parallelization algorithms can contribute to a great reduction in the time of execution and make high-fidelity turbulence simulation a more feasible option for predictive purposes.

One crucial area of future research is the construction of hybrid models employing reduced physics approximations where necessary. While the most accurate results can be obtained from full gyrokinetic simulations, their computational cost may be so high as to be unaffordable on long timescales for transport. By combining reduced models that capture the key physics but reduce the computational intensity, a compromise between the level of detail and efficiency might be found. These hybrids would make gyrokinetic turbulence modeling feasible in predictive contexts at a reasonable computational cost.

The communication between CPU and GPU elements of TRANSP also has areas for improvement. Optimizing load distribution between the two architectures is essential in precluding bottlenecks and achieving maximum utilization of resources. Optimizing data transfer algorithms and computational workload between CPU transport solvers and GPU turbulent calculations can enhance total efficiency and make it possible for simulations to be scalable and faster.

On top of the optimization of algorithms and hardware, optimizing numerical techniques in TRANSP can also make a difference in computational efficiency. Implementing adaptive time-stepping would enable the time step in the simulation to be varied dynamically according to plasma profile evolution and intensity of the turbulence. This would optimize computational effort by applying finer time resolution only when required while maintaining stability and accuracy throughout the simulation.

Together, these improvements would move TRANSP/T3D/GX calculations closer to practical runtimes for predictive and between-shot applications. By overcoming both the algorithms and the hardware hurdles, future advancements will facilitate better integration of high-accuracy turbulence modeling in everyday plasma transport simulations and improve the predictive capacity of TRANSP in the long run.

\section{Summary and Conclusions}
This paper describes two predictive numerical frameworks available in TRANSP. This description complements the earlier paper~\cite{APankin2024} that documented the TRANSP capabilities, including the predictive solver inputs and initializations, but did not describe the details of the algorithms of PT\_SOLVER or provide any verification. The predictive frameworks in TRANSP include the native PT\_SOLVER and the recently integrated T3D solver. PT\_SOLVER is designed as a parallel, modular, multi-region time-dependent transport solver, and it evolves flux surface average quantities using an implicit Newton-based technique. A set of six coupled equations for density, temperatures of electrons and ions, and toroidal angular momentum is evolved. The integration of transport equations includes various source couplings, geometrical motion terms, a nonlinear convergence constraint, and stabilization mechanisms using modified P\'{e}clet numbers to reduce stiffness related to gradient-sensitive closure relations. While T3D also solves coupled 1D equations and uses similar Newton iterations to evolve profiles in a time-dependent fashion, it is geared towards use with high-fidelity gyrokinetic simulations such as those provided by the GX code.

The numerical behavior of the PT\_SOLVER is evaluated through a series of hierarchical verification tests, including testing through both analytic and manufactured solutions, performing numerical convergence studies that included coupled equations with stiff gradient-dependent diffusion, and performing code-to-code verification against TGYRO with  TGLF and NEO transport models. Overall, these evaluations provide evidence that PT\_SOLVER achieves robust nonlinear convergence as well as profile solutions that are consistent with an established reference workflow, so long as equilibrium, sources, and boundary conditions are closely matched. In addition, the multilevel parallelization method, established through adopting the decomposition of computational domains over flux surfaces and providing for a high degree of internal parallelism in the transport model computations, enhances the practical applicability of PT\_SOLVER to performing computationally intensive predictive calculations.

This manuscript explains how the interface between TRANSP and the T3D transport solver works. The benchmarks for the T3D and PT\_SOLVER provide confirmation that the interface is correct and that the process flow is sound, but not a validation of the models for neoclassical and turbulent transport. As details of the T3D numerical algorithm are discussed elsewhere~\cite{Qian22,Barnes2008,Barnes2010}, they will not be repeated here. In the benchmark cases considered here, the coupled interface yields nearly identical profiles when T3D and PT\_SOLVER are supplied with the same source terms, impurity constraints, and equilibria. Minor differences arise from the different implementations of the collisional electron-ion exchange terms in PT\_SOLVER and T3D.

The T3D/TRANSP interface provides a new framework for predictive analysis with high-fidelity turbulence models and significantly enhances the fidelity of predictive computations in TRANSP. The framework can also be used to port the experiment profiles for linear stability analysis with GX. This additional benefit comes because TRANSP shares the interpretive and predictive workflows within the same framework as described in the TRANSP reference paper~\cite{APankin2024}.

The present work establishes the TRANSP predictive framework as a robust numerical component for predictive modeling and clarifies the present status of modular coupling to external transport and turbulence tools. Future work will focus on broader predictive applications, additional verification of extended transport channels, such as poloidal velocity advances, hybrid reduced and high-fidelity modeling strategies, and further optimization of GPU-enabled turbulence workflows.

\section*{Acknowledgments}

The research described in this paper was conducted at Princeton Plasma Physics Laboratory, a national laboratory operated by Princeton University for the United States Department of Energy under Prime Contract No. DE-AC02-09CH11466. {\ap{The PT\_SOLVER verification use the experimental data collected on the DIII-D tokamak at General Atomics supported by the U.S. Department of Energy,
Office of Science, Office of Fusion Energy Sciences, using the DIII-D
National Fusion Facility, a DOE Office of Science user facility, under
Award No. DE-FC02-04ER54698.}} The United States Government retains a non-exclusive, paid-up, irrevocable, worldwide license to publish or reproduce the published form of this manuscript, or allow others to do so, for United States Government purposes. This research used the resources of the National Energy Research Scientific Computing Center, a DOE Office of Science User Facility supported by the Office of Science of the U.S. Department of Energy under Contract No. DE-AC02-05CH11231.

The authors gratefully acknowledge the late Doug McCune for his leadership, sustained guidance, and encouragement over many years in the development of TRANSP and its predictive capabilities.

\subsection*{Data Availability}
\label{sec:data-repo}
Datasets and Python scripts associated with this work are deposited in the Princeton Data Commons (PDC) repository. The repository provides a persistent identifier (DOI) for the deposited materials. The data are released under the Creative Commons Attribution 4.0 International license (CC BY 4.0).\footnote{\url{https://creativecommons.org/licenses/by/4.0/}}\\
Repository landing page: 
\url{https://doi.org/10.34770/62xv-ra97}\\
DOI: 10.34770/62xv-ra97 \\
Users of these datasets should cite (i) this Article and (ii) the corresponding PDC dataset DOI.

\bibliographystyle{elsarticle-num}
\bibliography{refs}

@article{jardin08,
title = {On 1D diffusion problems with a gradient-dependent diffusion coefficient},
journal = {Journal of Computational Physics},
volume = {227},
number = {20},
pages = {8769-8775},
year = {2008},
issn = {0021-9991},
doi = {https://doi.org/10.1016/j.jcp.2008.06.032},
url = {https://www.sciencedirect.com/science/article/pii/S0021999108003616},
author = {S.C. Jardin and G. Bateman and G.W. Hammett and L.P. Ku}
}

@article{RHawryluk1980,
  author = {R.J. Hawryluk},
  title = {An Empirical Approach to Tokamak Transport},
  journal = {Physics of Plasmas Close to Thermonuclear Conditions, ed. by B. Coppi, et al.},
  publisher = {CEC, Brussels},
  year = {1980},
  volume = {Vol. 1},
  pages = {19-46}
}

@misc{JBreslau2018,
  author       = {Breslau, Joshua and Gorelenkova, Marina and Poli, Francesca and Sachdev, Jai and Pankin, Alexei and Perumpilly, Gopan and Yuan, Xingqiu and Glant, Laszlo},
  title        = {TRANSP},
  doi          = {10.11578/dc.20180627.4},
  url          = {https://www.osti.gov/biblio/code-12542},
  place        = {United States},
  year         = {2018},
  month        = {06}
  }

@article{Abbate2024,
  author  = {Abbate, J. and Fable, E. and Grierson, B. and Pankin, A. and Tardini, G. and Kolemen, E.},
  title   = {Large-database cross-verification and validation of tokamak transport models using baselines for comparison},
  journal = {Physics of Plasmas},
  year    = {2024},
  volume  = {31},
  number  = {4},
  pages   = {042506},
  doi     = {10.1063/5.0190908}
}

@article{Pereverzev2008,
title = {Stable numeric scheme for diffusion equation with a stiff transport},
journal = {Computer Physics Communications},
volume = {179},
number = {8},
pages = {579-585},
year = {2008},
issn = {0010-4655},
doi     = {10.1016/j.cpc.2008.05.006},
url = {https://www.sciencedirect.com/science/article/pii/S001046550800218X},
author = {G.V. Pereverzev and G. Corrigan}
}

@book{Jardin2010,
  title={Computational Methods in Plasma Physics},
  author={Jardin, Stephen},
  year={2010},
  publisher={CRC Press},
  address={Boca Raton},
  isbn={9780429075537},
  url={https://www.taylorfrancis.com/books/mono/10.1201/EBK1439810958/computational-methods-plasma-physics-stephen-jardin},
  doi       = {10.1201/EBK1439810958}
}

@article{VanMulders2021,
  author  = {Van Mulders, Simon and Felici, Federico and Sauter, O. and Citrin, J. and Ho, A. and Marin, M. and van de Plassche, K. L.},
  title   = {Rapid optimization of stationary tokamak plasmas in {RAPTOR}: demonstration for the {ITER} hybrid scenario with neural network surrogate transport model {QLKNN}},
  journal = {Nuclear Fusion},
  year    = {2021},
  volume  = {61},
  number  = {8},
  pages   = {086019},
  doi     = {10.1088/1741-4326/ac0d12}
}

@article{Candy2009,
  author  = {Candy, J. and Holland, C. and Waltz, R. E. and Fahey, M. R. and Belli, E. A.},
  title   = {Tokamak profile prediction using direct gyrokinetic and neoclassical simulation},
  journal = {Physics of Plasmas},
  year    = {2009},
  volume  = {16},
  number  = {6},
  pages   = {060704},
  doi     = {10.1063/1.3167820}
}

@article{Barnes2010,
  author  = {Barnes, M. and Abel, I. G. and Dorland, W. and G{\"o}rler, T. and Hammett, G. W. and Jenko, F.},
  title   = {Direct multiscale coupling of a transport code to gyrokinetic turbulence codes},
  journal = {Physics of Plasmas},
  year    = {2010},
  volume  = {17},
  number  = {5},
  pages   = {056109},
  doi     = {10.1063/1.3323082}
}

@article{Jardin1986,
  author  = {Jardin, S. C. and Pomphrey, N. and DeLucia, J.},
  title   = {Dynamic modeling of transport and positional control of tokamaks},
  journal = {Journal of Computational Physics},
  year    = {1986},
  volume  = {66},
  number  = {2},
  pages   = {481--507},
  doi     = {10.1016/0021-9991(86)90077-X}
}

@article{Park2017,
  author  = {Park, J. M. and Murakami, M. and St. John, H. E. and Lao, L. L. and Chu, M. S. and Prater, R.},
  title   = {An efficient transport solver for tokamak plasmas},
  journal = {Computer Physics Communications},
  year    = {2017},
  volume  = {214},
  pages   = {1--5},
  doi     = {10.1016/j.cpc.2016.12.018}
}

@article{LudvigOsipov2025,
  author  = {Ludvig-Osipov, Andrei and Yadykin, Dmytro and Strand, P{\"a}r},
  title   = {High-order implicit solver in conservative formulation for tokamak plasma transport equations},
  journal = {Computer Physics Communications},
  year    = {2025},
  volume  = {311},
  pages   = {109570},
  doi     = {10.1016/j.cpc.2025.109570}
}

@article{Felici2018,
  author  = {Felici, Federico and Citrin, J. and Teplukhina, A. A. and Redondo, J. and Bourdelle, C. and Imbeaux, F. and Sauter, O.},
  title   = {Real-time-capable prediction of temperature and density profiles in a tokamak using {RAPTOR} and a first-principle-based transport model},
  journal = {Nuclear Fusion},
  year    = {2018},
  volume  = {58},
  number  = {9},
  pages   = {096006},
  doi     = {10.1088/1741-4326/aac8f0}
}

@article{Mandell2018,
  author = {Mandell, N. R. and Dorland, W. and Landreman, M.},
  title = {Laguerre-Hermite pseudo-spectral velocity formulation of gyrokinetics},
  journal = {Journal of Plasma Physics},
  volume = {84},
  number = {1},
  pages = {905840108},
  year = {2018},
  doi = {10.1017/S0022377818000041}
}

@article{Mandell2024,
  author = {Mandell, N. R. and Dorland, W. and Abel, I. and Gaur, R. and Kim, P. and Martin, M. and Qian, T.},
  title = {GX: a GPU-native gyrokinetic turbulence code for tokamak and stellarator design},
  journal = {Journal of Plasma Physics},
  volume = {90},
  number = {4},
  pages = {905900402},
  year = {2024},
  doi = {10.1017/S0022377824000631}
}

@article{Giacomin2025,
  author = {Giacomin, M. and Dickinson, D. and Dorland, W. and Mandell, N. R. and Bokshi, A. and Casson, F. J. and Dudding, H. G. and Kennedy, D. and Patel, B. S. and Roach, C. M.},
  title = {A quasi-linear model of electromagnetic turbulent transport and its application to flux-driven transport predictions for STEP},
  journal = {Journal of Plasma Physics},
  volume = {91},
  number = {1},
  pages = {E16},
  year = {2025},
  doi = {10.1017/S0022377824001107}
}

@misc{transpT3DGXGuide,
  author = {{Princeton Plasma Physics Laboratory}},
  title = {Guide to Running TRANSP/T3D/GX Coupled Simulations},
  year = {2026},
  note = {TRANSP documentation page},
  url = {https://transp.pppl.gov/modules/t3d.html}
}

@INPROCEEDINGS{Qian22,
       author = {{Qian}, Tony and {Buck}, Braden and {Gaur}, Rahul and {Mandell}, Noah and {Kim}, Patrick and {Dorland}, William},
    title = {Stellarator profile predictions using {Trinity3D} and {GX}},
    booktitle = {APS Division of Plasma Physics Meeting Abstracts},
         year = 2022,
       series = {APS Meeting Abstracts},
       volume = {2022},
        month = jan,
        pages = {BO03.006},
       adsurl = {https://ui.adsabs.harvard.edu/abs/2022APS..DPPBO3006Q}
}

@phdthesis{Barnes2008,
  author       = {Barnes, Michael},
  title        = {{TRINITY}: A unified treatment of turbulence, transport, and heating in magnetized plasmas},
  school       = {University of Maryland},
  year         = {2008},
  type         = {Ph.D. thesis},
  note         = {eprint arXiv:0901.2868},
  eprint       = {0901.2868},
  archivePrefix= {arXiv},
  primaryClass = {physics.plasm-ph}
}

@article{Auriemma_2026,
doi = {10.1088/1741-4326/ae5de2},
url = {https://doi.org/10.1088/1741-4326/ae5de2},
year = {2026},
month = {apr},
publisher = {IOP Publishing},
volume = {66},
number = {5},
pages = {056043},
author = {Auriemma, F. and Kim, Hyun-Tae and Lombardo, J. and Lorenzini, R. and Gabriellini, S. and Garzotti, L. and Zotta, V.K. and Frigione, D. and Menmuir, S. and Gorelenkova, M. and Pankin, A.Y. and Rimini, F. and Staebler, G.M. and \v{S}tancar, \v{Z}. and Van Eester, D. and Garcia, J. and Hillesheim, J. and Lomas, P. and Lowry, C. and JET Contributors and the EUROfusion Tokamak Exploitation Team},
title = {{TRANSP-TGLF} core predictive modeling of the {JET} {DT} baseline scenario},
journal = {Nuclear Fusion}
}

@article{Kim_2023,
doi = {10.1088/1741-4326/ace26d},
url = {https://doi.org/10.1088/1741-4326/ace26d},
year = {2023},
month = {oct},
publisher = {IOP Publishing},
volume = {63},
number = {11},
pages = {112004},
author = {Kim, Hyun-Tae and Auriemma, Fulvio and Ferreira, Jorge and Gabriellini, Stefano and Ho, Aaron and Huynh, Philippe and Kirov, Krassimir and Lorenzini, Rita and Marin, Michele and Poradzinski, Michal and Shi, Nan and Staebler, Gary and \v{S}tancar, \v{Z}iga and Stankunas, Gediminas and Konrad Zotta, Vito and Belli, Emily and Casson, Francis J and D Challis, Clive and Citrin, Jonathan and van Eester, Dirk and Fransson, Emil and Gallart, Daniel and Garcia, Jeronimo and Garzotti, Luca and Gatto, Renato and Hobirk, Joerg and Kappatou, Athina and Lerche, Ernesto and Ludvig-Osipov, Andrei and Maggi, Costanza and Maslov, Mikhail and Nocente, Massimo and Sharma, Ridhima and Di Siena, Alessandro and Strand, Par and Tholerus, Emmi and Yadykin, Dimitriy and JET Contributors},
title = {Validation of {D–T} fusion power prediction capability against 2021 {JET} {D–T} experiments},
journal = {Nuclear Fusion}
}

@article{Lestz2025_AssessingNSTX,
  author  = {Lestz, J. B. and Avdeeva, G. and Neiser, T. F. and Gorelenkova, M. V. and Halpern, F. D. and Kaye, S. M. and McClenaghan, J. and Pankin, A. Y. and Thome, K. E.},
  title   = {Assessing time-dependent temperature profile predictions using reduced transport models for high performing NSTX plasmas},
  journal = {Plasma Physics and Controlled Fusion},
  volume  = {67},
  number  = {10},
  pages   = {105029},
  year    = {2025},
  doi     = {10.1088/1361-6587/ae0c35}
}

@article{Lestz2025_MMMSensitivityNSTX,
  author  = {Lestz, J. B. and Avdeeva, G. and Kaye, S. M. and Gorelenkova, M. V. and Halpern, F. D. and McClenaghan, J. and Pankin, A. Y. and Thome, K. E.},
  title   = {Sensitivities of time-dependent temperature profile predictions for NSTX with the Multi-Mode Model},
  journal = {Plasma Physics and Controlled Fusion},
  volume  = {67},
  number  = {10},
  pages   = {105030},
  year    = {2025},
  doi     = {10.1088/1361-6587/ae0c34}
}

@article{Avdeeva_2023,
doi = {10.1088/1741-4326/acfc56},
url = {https://doi.org/10.1088/1741-4326/acfc56},
year = {2023},
month = {oct},
publisher = {IOP Publishing},
volume = {63},
number = {12},
pages = {126020},
author = {Avdeeva, G. and Thome, K.E. and Smith, S.P. and Battaglia, D.J. and Clauser, C.F. and Guttenfelder, W. and Kaye, S.M. and McClenaghan, J. and Meneghini, O. and Odstrcil, T. and Staebler, G.},
title = {Energy transport analysis of {NSTX} plasmas with the {TGLF} turbulent and {NEO} neoclassical transport models},
journal = {Nuclear Fusion}
}

@article{Budny_2012,
doi = {10.1088/0029-5515/52/1/013001},
url = {https://doi.org/10.1088/0029-5515/52/1/013001},
year = {2011},
month = {nov},
publisher = {},
volume = {52},
number = {1},
pages = {013001},
author = {Budny, R.V.},
title = {Alpha heating in {ITER} {L}-mode and {H}-mode plasmas},
journal = {Nuclear Fusion}
}

@article{BudnyCordey2016,
  author  = {Budny, R. V. and Cordey, J. G.},
  title   = {Core fusion power gain and alpha heating in JET, TFTR, and ITER},
  journal = {Nuclear Fusion},
  volume  = {56},
  pages   = {056002},
  year    = {2016}
}

@article{BudnyJET2018,
  author  = {Budny, R. V. and JET Contributors},
  title   = {Alpha heating, isotopic mass, and fast ion effects in deuterium--tritium experiments},
  journal = {Nuclear Fusion},
  volume  = {58},
  pages   = {096011},
  year    = {2018}
}

@article{APankin2024,
title = {TRANSP integrated modeling code for interpretive and predictive analysis of tokamak plasmas},
journal = {Computer Physics Communications},
volume = {312},
pages = {109611},
year = {2025},
issn = {0010-4655},
doi = {https://doi.org/10.1016/j.cpc.2025.109611},
url = {https://www.sciencedirect.com/science/article/pii/S0010465525001134},
author = {A.Y. Pankin and J. Breslau and M. Gorelenkova and R. Andre and B. Grierson and J. Sachdev and M. Goliyad and G. Perumpilly}
}

@article{Staebler2005TGLF,
  author  = {Staebler, G. M. and Kinsey, J. E. and Waltz, R. E.},
  title   = {Gyro-Landau fluid equations for trapped and passing particles},
  journal = {Physics of Plasmas},
  volume  = {12},
  number  = {10},
  pages   = {102508},
  year    = {2005},
  doi     = {10.1063/1.2044587}
}

@article{Staebler2007TGLF,
  author  = {Staebler, G. M. and Kinsey, J. E. and Waltz, R. E.},
  title   = {A theory-based transport model with comprehensive physics},
  journal = {Physics of Plasmas},
  volume  = {14},
  number  = {5},
  pages   = {055909},
  year    = {2007},
  doi     = {10.1063/1.2436852}
}

@article{Kinsey2008TGLF,
  author  = {Kinsey, J. E. and Staebler, G. M. and Waltz, R. E.},
  title   = {The first transport code simulations using the trapped gyro-Landau-fluid model},
  journal = {Physics of Plasmas},
  volume  = {15},
  number  = {5},
  pages   = {055908},
  year    = {2008},
  doi     = {10.1063/1.2889008}
}

@article{Belli2008KineticCO,
  title={Kinetic calculation of neoclassical transport including self-consistent electron and impurity dynamics},
  author={Emily Ann Belli and Jeff Candy},
  journal={Plasma Physics and Controlled Fusion},
  year={2008},
  volume={50},
  pages={095010}
}

@article{GLF23_1,
author = {Waltz,R. E.  and Staebler,G. M.  and Dorland,W.  and Hammett,G. W.  and Kotschenreuther,M.  and Konings,J. A. },
title = {A gyro-Landau-fluid transport model},
journal = {Physics of Plasmas},
volume = {4},
number = {7},
pages = {2482-2496},
year = {1997},
doi = {10.1063/1.872228},
URL = {         https://doi.org/10.1063/1.872228}}

@article{GLF23_2,
author = {Jonathan E. Kinsey and Gary M. Staebler and Ronald E. Waltz},
title = {Burning Plasma Confinement Projections and Renormalization of the GLF23 Drift-Wave Transport Model},
journal = {Fusion Science and Technology},
volume = {44},
number = {4},
pages = {763-775},
year  = {2003},
publisher = {Taylor & Francis},
doi = {10.13182/FST03-A414},
URL = {         https://doi.org/10.13182/FST03-A414    }}

@article{Rafiq:pop20,
title = {Physics basis of Multi-Mode anomalous transport module},
author = {T. Rafiq and A. H. Kritz and J. Weiland and A. Y. Pankin and L. Luo},
journal = {Physics of Plasmas},
volume = {20},
number = {3},
pages = {032506},
year = {2013},
DOI={10.1063/1.4794288}
}

@article{Rafiq:pop23,
title = {Microtearing modes in tokamak discharges},
author = {T. Rafiq and J. Weiland and A. H. Kritz and L. Luo and A. Y. Pankin},
journal = {Physics of Plasmas},
volume = {23},
number = {6},
pages = {062507},
year = {2016},
DOI={10.1063/1.4953609}
}

@article{TGLF,
  title={Gyro-{Landau} fluid equations for trapped and passing particles},
  author={G. M. Staebler and J. E. Kinsey and R. E. Waltz },
  journal={Phys. Plasmas},
  volume={12},
  pages={102508},
  year={2005},
  DOI={10.1063/1.2044587}
}

@techreport{Pereverzev2002_ASTRA,
  author      = {Pereverzev, G. and Yushmanov, P.},
  title       = {{ASTRA} automated system for TRansport analysis in a tokamak},
  institution = {Max-Planck-Institut f{\"u}r Plasmaphysik},
  number      = {IPP 5/98},
  year        = {2002}
}

@article{Park2018_IPSFASTRAN,
  author  = {Park, J. M. and Ferron, J. R. and Holcomb, C. T. and Buttery, R. J. and Solomon, W. M. and Batchelor, D. B. and Elwasif, W. and Green, D. L. and Kim, K. and Meneghini, O. and others},
  title   = {Integrated modeling of high $\beta_N$ steady state scenario on {DIII-D}},
  journal = {Physics of Plasmas},
  volume  = {25},
  pages   = {012506},
  year    = {2018}
}

@article{Romanelli2014_JINTRAC,
  author  = {Romanelli, M. and Corrigan, G. and Parail, V. and Wiesen, S. and Ambrosino, R. and da Silva Aresta Belo, P. and Garzotti, L. and Harting, D. and K{\"o}chl, F. and Koskela, T. and Lauro-Taroni, L. and Marchetto, C. and Mattei, M. and Militello-Asp, E. and Nave, M. F. F. and Pamela, S. and Salmi, A. and Strand, P. and Szepesi, G. and EFDA-JET Contributors},
  title   = {{JINTRAC}: A system of codes for integrated simulation of Tokamak scenarios},
  journal = {Plasma and Fusion Research},
  volume  = {9},
  pages   = {3403023},
  year    = {2014},
  doi     = {10.1585/pfr.9.3403023}
}

@techreport{Cenacchi1988_JETTO,
  author      = {Cenacchi, G. and Taroni, A.},
  title       = {{JETTO}: A Free Boundary Plasma Transport Code (Basic Version)},
  institution = {JET / ENEA},
  number      = {JET-IR(88)03},
  year        = {1988}
}

@article{Meneghini2024_FUSE,
  author  = {Meneghini, O. and Slendebroek, T. and Lyons, B. C. and McLaughlin, K. and McClenaghan, J. and Stagner, L. and Harvey, J. and Neiser, T. F. and Ghiozzi, A. and Dose, G. and Guterl, J. and Zalzali, A. and Cote, T. and Shi, N. and Weisberg, D. and Smith, S. P. and Grierson, B. A. and Candy, J.},
  title   = {{FUSE} (Fusion Synthesis Engine): A Next Generation Framework for Integrated Design of Fusion Pilot Plants},
  journal = {arXiv preprint arXiv:2409.05894},
  year    = {2024},
  doi     = {10.48550/arXiv.2409.05894},
  eprint  = {2409.05894},
  archivePrefix = {arXiv},
  primaryClass  = {physics.plasm-ph}
}

@article{Citrin2024_TORAX,
  author  = {Citrin, Jonathan and Goodfellow, Ian and Raju, Akhil and Chen, Jeremy and Degrave, Jonas and Donner, Craig and Felici, Federico and Hamel, Philippe and Huber, Andrea and Nikulin, Dmitry and Pfau, David and Tracey, Brendan and Riedmiller, Martin and Kohli, Pushmeet},
  title   = {{TORAX}: A Fast and Differentiable Tokamak Transport Simulator in {JAX}},
  journal = {arXiv preprint arXiv:2406.06718},
  year    = {2024},
  doi     = {10.48550/arXiv.2406.06718},
  eprint  = {2406.06718},
  archivePrefix = {arXiv},
  primaryClass  = {physics.plasm-ph}
}

@inproceedings{rebut89,
  author={Rebut, P-H. and Lallia, P.P. and Watkins, M.L.},
  booktitle={Plasma Physics and Controlled Nuclear Fusion Research (Proc. 12th IAEA Conf., Nice, France)},
  title={The critical temperature gradient model of plasma transport: Applications to {JET} and future tokamaks},
  volume=2, 
  publisher={IAEA}, 
  year=1989,
  pages=191
  }

@article{jardin93,
  doi = {10.1088/0029-5515/33/3/I01},
  url = {https://dx.doi.org/10.1088/0029-5515/33/3/I01},
  year = {1993},
  month = {mar},
  publisher = {},
  volume = {33},
  number = {3},
  pages = {371},
  author = {S.C. Jardin and  M.G. Bell and  N. Pomphrey},
  title = {{TSC} simulation of Ohmic discharges in {TFTR}}, 
  journal = {Nuclear Fusion},
  }

@article{fukuyama95,
  doi = {10.1088/0741-3335/37/6/002},
  url = {https://dx.doi.org/10.1088/0741-3335/37/6/002},
  year = {1995},
  month = {jun},
  publisher = {},
  volume = {37},
  number = {6},
  pages = {611},
  author = {A Fukuyama and  K Itoh and  S -I Itoh and  M Yagi and  M Azumi},
  title = {Transport simulation on {L-mode} and improved confinement associated with current profile modification},
  journal = {Plasma Physics and Controlled Fusion},
  }

@article{callen05,
  doi = {10.1088/0029-5515/45/9/012},
  url = {https://dx.doi.org/10.1088/0029-5515/45/9/012},
  year = {2005},
  month = {aug},
  publisher = {},
  volume = {45},
  number = {9},
  pages = {1120},
  author = {J.D. Callen},
  title = {Paleoclassical electron heat transport},
  journal = {Nuclear Fusion},
  }

@article{takei07,
  doi = {10.1088/0741-3335/49/3/011},
  url = {https://dx.doi.org/10.1088/0741-3335/49/3/011},
  year = {2007},
  month = {feb},
  publisher = {},
  volume = {49},
  number = {3},
  pages = {335},
  author = {N Takei and Y Nakamura and M Ushigome and T Suzuki and N Aiba and M Takechi and K Tobita and Y Takase and A Fukuyama and S C Jardin},
  title = {Intermittent $\beta$ collapse after {NBCD} turn-off in {JT-60U} fully non-inductive reversed shear discharges},
  journal = {Plasma Physics and Controlled Fusion},
  }

@article{belli12,
  doi = {10.1088/0741-3335/54/1/015015},
  url = {https://dx.doi.org/10.1088/0741-3335/54/1/015015},
  year = {2011},
  month = {dec},
  publisher = {},
  volume = {54},
  number = {1},
  pages = {015015},
  author = {E A Belli and J Candy},
  title = {Full linearized Fokker–Planck collisions in neoclassical transport simulations},
  journal = {Plasma Physics and Controlled Fusion}
  }

@article{belli08,
  doi = {10.1088/0741-3335/50/9/095010},
  url = {https://dx.doi.org/10.1088/0741-3335/50/9/095010},
  year = {2008},
  month = {jul},
  publisher = {},
  volume = {50},
  number = {9},
  pages = {095010},
  author = {E A Belli and J Candy},
  title = {Kinetic calculation of neoclassical transport including self-consistent electron and impurity dynamics},
  journal = {Plasma Physics and Controlled Fusion}
  }

@article{houlber97,
      author = {Houlberg, W. A. and Shaing, K. C. and Hirshman, S. P. and Zarnstorff, M. C.},
      title = {Bootstrap current and neoclassical transport in tokamaks of arbitrary collisionality and aspect ratio},
      journal = {Physics of Plasmas},
      volume = {4},
      number = {9},
      pages = {3230-3242},
      year = {1997},
      month = {09},
      issn = {1070-664X},
      doi = {10.1063/1.872465},
      url = {https://doi.org/10.1063/1.872465},
      eprint = {https://pubs.aip.org/aip/pop/article-pdf/4/9/3230/19108363/3230\_1\_online.pdf},
  }

@article{Dudding_2022,
	author = {Dudding, H.G. and Casson, F.J. and Dickinson, D. and Patel, B.S. and Roach, C.M. and Belli, E.A. and Staebler, G.M.},
	doi = {10.1088/1741-4326/ac7a4d},
	journal = {Nuclear Fusion},
	month = {jul},
	number = {9},
	pages = {096005},
	publisher = {IOP Publishing},
	title = {A new quasilinear saturation rule for tokamak turbulence with application to the isotope scaling of transport},
	url = {https://doi.org/10.1088/1741-4326/ac7a4d},
	volume = {62},
	year = {2022}
}

@inproceedings{Neiser2022_TGLFNNDatabase,
  author    = {Neiser, Tom F. and Meneghini, Orso and Smith, Sterling P. and McClenaghan, Joseph T. and Orozco, David and Hall, Joseph B. and Staebler, Gary M. and Belli, Emily A. and Candy, Jeff},
  title     = {Database generation for validation of TGLF and retraining of neural network accelerated {TGLF-NN}},
  booktitle = {APS Division of Plasma Physics Meeting Abstracts},
  year      = {2022},
  pages     = {GP11.00010},
  note      = {Abstract GP11.00010}
}

@inproceedings{Neiser2023_MultifidelityNN,
  author    = {Neiser, Tom F. and Meneghini, Orso and Smith, Sterling P. and McClenaghan, Joseph T. and Slendebroek, Tim and Orozco, David and Sammuli, Brian and Staebler, Gary M. and Hall, Joseph B. and Belli, Emily A. and Candy, Jeff},
  title     = {Multi-fidelity neural network representation of gyrokinetic turbulence},
  booktitle = {APS Division of Plasma Physics Meeting Abstracts},
  year      = {2023},
  pages     = {PP11.00039},
  note      = {Abstract PP11.00039}
}

@article{chang86,
      author = {Chang, C. S. and Hinton, F. L.},
      title = {Effect of impurity particles on the finite‐aspect ratio neoclassical ion thermal conductivity in a tokamak},
      journal = {The Physics of Fluids},
      volume = {29}, 
      number = {10},
      pages = {3314-3316},
      year = {1986},
      month = {10},
      issn = {0031-9171},
      doi = {10.1063/1.865847},
      url = {https://doi.org/10.1063/1.865847},
      eprint = {https://pubs.aip.org/aip/pfl/article-pdf/29/10/3314/12413384/3314\_1\_online.pdf}
}

@conference{Neiser2024TTF,
author = {Neiser, T.F.},
title = {Large Database Validation of {TGLF} on {DIII-D} and {MAST-U}
Plasmas},
booktitle = {US-EU Transport Task Force},
year = {2024},
address = {Asheville, NC},
month = {Apr.~9-12},
url = {https://conferences.union.wisc.edu/ttf/conference-agenda/}
}

@article{Kinsey2011,
doi = {10.1088/0029-5515/51/8/083001},
url = {https://doi.org/10.1088/0029-5515/51/8/083001},
year = {2011},
month = {jun},
publisher = {},
volume = {51},
number = {8},
pages = {083001},
author = {Kinsey, J.E. and Staebler, G.M. and Candy, J. and Waltz, R.E. and Budny, R.V.},
title = {{ITER} predictions using the {GYRO} verified and experimentally validated trapped gyro-{Landau} fluid transport model},
journal = {Nuclear Fusion}
}

@article{Wang2010,
    author = {Wang, W. X. and Diamond, P. H. and Hahm, T. S. and Ethier, S. and Rewoldt, G. and Tang, W. M.},
    title = {Nonlinear flow generation by electrostatic turbulence in tokamaks},
    journal = {Physics of Plasmas},
    volume = {17},
    number = {7},
    pages = {072511},
    year = {2010},
    month = {07},
    issn = {1070-664X},
    doi = {10.1063/1.3459096},
    url = {https://doi.org/10.1063/1.3459096},
    eprint = {https://pubs.aip.org/aip/pop/article-pdf/doi/10.1063/1.3459096/15641806/072511_1_online.pdf},
}

@article{Horton1999,
doi = {10.1088/0029-5515/39/8/304},
url = {https://doi.org/10.1088/0029-5515/39/8/304},
year = {1999},
month = {aug},
publisher = {},
volume = {39},
number = {8},
pages = {993},
author = {L.D. Horton and R. Sartori and B. Balet and R.V. Budny and J.P. Christiansen and S. Clement and G.D. Conway and J.G. Cordey and G.M. Fishpool and J. Lingertat and C.G. Lowry and C.F. Maggi and M.J. Mantsinen and V. Riccardo and G. Saibene and P. Smeulders and R.J. Smith and K. Thomsen and M.G. von Hellermann},
title = {High fusion power steady state operation in {JET} {DT} plasmas},
journal = {Nuclear Fusion}
}

@article{Zotta2022,
doi = {10.1088/1741-4326/ac5f19},
url = {https://doi.org/10.1088/1741-4326/ac5f19},
year = {2022},
month = {may},
publisher = {IOP Publishing},
volume = {62},
number = {7},
pages = {076024},
author = {Zotta, V.K. and Garzotti, L. and Casson, F.J. and Frigione, D. and Köchl, F. and Lerche, E. and Lomas, P. and Rimini, F. and Sertoli, M. and Van Eester, D. and Gatto, R. and Mazzotta, C. and Pucella, G. and JET Contributors},
title = {Fusion power predictions for $\beta_N \approx 1.8$ baseline scenario with 50–50 {D–T} fuel mix and {NBI} injection in preparation to {D–T} operations at {JET}},
journal = {Nuclear Fusion}
}

@article{Solomon2007,
  author  = {Solomon, W. M. and Burrell, K. H. and DeGrassie, J. S. and Budny, R. and Groebner, R. J. and Kinsey, J. E. and Kramer, G. J. and Luce, T. C. and Makowski, M. A. and Mikkelsen, D. and Nazikian, R. and Petty, C. C. and Politzer, P. A. and Scott, S. D. and {Van Zeeland}, M. A. and Zarnstorff, M. C.},
  title   = {Momentum confinement at low torque},
  journal = {Plasma Physics and Controlled Fusion},
  volume  = {49},
  number  = {12B},
  pages   = {B313--B324},
  year    = {2007},
  doi     = {10.1088/0741-3335/49/12B/S29}
}

@inproceedings{Budny2012,
  author       = {Budny, R. V. and Yuan, Xingqiu and Jardin, S. and Hammett, G. and Staebler, G. and Kinsey, J. and {members of the ITPA Transport and Confinement Topical Group} and {JET EFDA Contributors}},
  title        = {{PTRANSP} Tests of {TGLF} and Predictions for {ITER}},
  booktitle    = {Proceedings of the 24th IAEA Fusion Energy Conference},
  eventtitle   = {24th IAEA Fusion Energy Conference},
  venue        = {San Diego, California, USA},
  year         = {2012},
  note         = {IAEA-CN-197, contribution ITR/P1-29}
}

@article{Budny2000,
  author  = {Budny, R. V. and Ernst, D. R. and Hahm, T. S. and McCune, D. C. and Christiansen, J. P. and Cordey, J. G. and Gowers, C. G. and Guenther, K. and Hawkes, N. and Jarvis, O. N. and Stubberfield, P. M. and Zastrow, K.-D. and Horton, L. D. and Saibene, G. and Sartori, R. and Thomsen, K. and {von Hellermann}, M. G.},
  title   = {Local transport in {Joint European Tokamak} edge-localized, high-confinement mode plasmas with {H}, {D}, {DT}, and {T} isotopes},
  journal = {Physics of Plasmas},
  volume  = {7},
  number  = {12},
  pages   = {5038--5050},
  year    = {2000},
  doi     = {10.1063/1.1320466}
}
\end{document}